\PassOptionsToPackage{pdfpagelabels=false}{hyperref} 
\documentclass{aa}

\usepackage{natbib}
\usepackage{graphicx}
\usepackage{txfonts}
\usepackage{amsmath}
\usepackage{amssymb}	% Extra maths symbols
\usepackage{rotating}
\usepackage{natbib}
\usepackage{gensymb}

\bibpunct{(}{)}{;}{a}{}{,} % to follow the A&A style
\usepackage[colorlinks,linkcolor=blue,anchorcolor=blue,citecolor=blue]{hyperref}

\graphicspath{{./}{Figures}}

\begin{document}

% \label{firstpage}
% \pagerange{\pageref{firstpage}--\pageref{lastpage}}

\title{Particle re-acceleration and diffuse radio sources in the galaxy cluster Abell 1550}

\author{T. Pasini\inst{\ref{inst1}}
\and H. W. Edler\inst{\ref{inst1}}
\and M. Brüggen\inst{\ref{inst1}}
\and F. de Gasperin\inst{\ref{inst1},\ref{inst2}}
\and A. Botteon\inst{\ref{inst3}}
\and K. Rajpurohit\inst{\ref{inst2}, \ref{inst4}, \ref{inst5}}
\and R. J. van Weeren\inst{\ref{inst3}}
\and F. Gastaldello\inst{\ref{inst6}}
\and M. Gaspari\inst{\ref{inst7}, \ref{inst8}}
\and G. Brunetti\inst{\ref{inst2}}
\and V. Cuciti\inst{\ref{inst1}}
\and C. Nanci\inst{\ref{inst2},\ref{inst4}}
\and G. di Gennaro\inst{\ref{inst1}}
\and M. Rossetti\inst{\ref{inst6}}
\and D. Dallacasa\inst{\ref{inst4}}
\and D. N. Hoang\inst{\ref{inst1}} 
\and C. J. Riseley\inst{\ref{inst4}, \ref{inst2}, \ref{inst9}}
}
\authorrunning{T. Pasini, H. W. Edler, M. Brüggen et al.}

%\author[0000-0003-2754-9258]{M.~Gaspari}

% List of institutions
\institute{Hamburger Sternwarte, Universität Hamburg, Gojenbergsweg 112, 21029 Hamburg, Germany\label{inst1}\\ \email{thomas.pasini@hs.uni-hamburg.de}
\and INAF - Istituto di Radioastronomia, via P. Gobetti 101, 40129, Bologna, Italy\label{inst2}
\and Leiden Observatory, Leiden University, PO Box 9513, NL-2300 RA Leiden, The Netherlands\label{inst3}
\and Dipartimento di Fisica e Astronomia, Universit\`a di Bologna, via Gobetti 93/2, 40129 Bologna, Italy\label{inst4}
\and Thüringer Landessternwarte, Sternwarte 5, 07778 Tautenburg, Germany\label{inst5}
\and INAF - IASF Milano, via A. Corti 12, I-20133 Milano, Italy\label{inst6}
\and INAF - Osservatorio di Astrofisica e Scienza dello Spazio, via P. Gobetti 93/3, I-40129 Bologna, Italy\label{inst7}
\and Department of Astrophysical Sciences, Princeton University, 4 Ivy Lane, Princeton, NJ 08544-1001, USA\label{inst8}
\and CSIRO Space \& Astronomy, PO Box 1130, Bentley, WA 6102, Australia\label{inst9}
}

%\date{Accepted 2020 July 8. Received 2020 July 8; in original form 2020 May 6}

%\pubyear{2022}

\abstract
{Radio observations of galaxy clusters reveal a plethora of diffuse, steep-spectrum sources related to the re-acceleration of cosmic-ray electrons, such as halos, relics and phoenices. In this context, the LOFAR LBA Sky Survey provides the most sensitive images of the sky at 54 MHz to date, allowing to investigate re-acceleration processes in a poorly-explored frequency regime.}
{We study diffuse radio emission in the galaxy cluster Abell 1550, with the aim of constraining particle re-acceleration in the intra-cluster medium.}
{We exploit observations at four different radio frequencies: 54, 144, 400 and 1400 MHz. To complement our analysis, we make use of archival \textit{Chandra} X-ray data.}
{At all frequencies we detect an ultra-steep spectrum radio halo ($S_\nu \propto \nu^{-1.6}$) with an extent of $\sim$ 1.2 Mpc at 54 MHz. Its morphology follows the distribution of the thermal intra-cluster medium inferred from the \textit{Chandra} observation. 
%In the northernmost region we detect an extension which seems to be part of the same halo. 
West of the centrally located head-tail radio galaxy, we detect a radio relic with a projected extent of $\sim$ 500 kpc. From the relic, a $\sim$600 kpc long bridge departs and connect it with the halo. Between the relic and the radio galaxy, we observe what is most likely a radio phoenix, given its curved spectrum. The phoenix is connected to the tail of the radio galaxy through two arms, which show a nearly constant spectral index for $\sim$ 300 kpc.}
{The halo could be produced by turbulence induced by a major merger, with the merger axis lying in the NE-SW direction. This is supported by the position of the relic, whose origin could be attributed to a shock propagating along the merger axis. It is possible that the same shock has also produced the phoenix through adiabatic compression, while we propose that the bridge could be generated by electrons which were pre-accelerated by the shock, and then re-accelerated by turbulence. Finally, we detect hints of gentle re-energisation in the two arms which depart from the tail of the radio galaxy.}


\maketitle

%%%%%%%%%%%%%%%%%%%%%%%%%%%%%%%%%%%%%%%%%%%%%%%%%%%%%%%%%%%%%%%%%%%%%%%%%%%%%%

\section{Introduction} 
\label{sec:intro}

Radio observations reveal the presence of diffuse synchrotron emission in galaxy clusters which is not directly associated with galaxies. This kind of emission is thought to be produced by the (re-)acceleration of cosmic-ray (CR) electrons due to shocks and turbulence in the Intra-Cluster Medium (ICM) \citep{Bruggen_2012, Brunetti_2014}. The synchrotron spectra of these electrons typically show a power-law distribution, $S(\nu) \propto \nu^{\alpha}$, with $S(\nu)$ being the flux density at frequency $\nu$, and $\alpha$ being the spectral index.

Giant radio halos (RH) are Mpc-scale sources centered in the central regions of merging clusters \citep{Cassano_2010}. They usually exhibit spherically symmetric morphology, even though filamentary structures are sometimes detected \citep{vanWeeren_2017, Botteon_2020a}. The radio emission is spatially correlated to the distribution of the ICM revealed by X-ray observations \citep{Govoni_2001, Giacintucci_2005, Rajpurohit_2018}, suggesting a link between thermal and non-thermal plasma. The integrated spectral index of RHs typically ranges between $-1.1 \leq \alpha \leq -1.4$ \citep{Giovannini_2009, Feretti_2012}. Nevertheless, an increasing number of RHs with ultra-steep spectra (USSRH) were discovered, with indices in the range $-1.5 \leq \alpha \leq -2$ \citep[e.g.][]{Brunetti_2008, Macario_2010, Dallacasa_2009, Bonafede_2012, Wilber_2018, Bruno_2021, Duchesne_2022}. The luminosity of RHs correlates with the host cluster's X-ray luminosity, making their detection easier in high-mass objects \citep[e.g.,][]{Cassano_2013, Cuciti_2015}. Radio halos are mainly observed in merging clusters \citep{Cassano_2013, Cuciti_2015} and their origin is traced back to mechanisms driven by large-scale turbulence \citep[e.g.,][]{Brunetti_2009, Cassano_2010, Eckert_2017}. \citet{Cuciti_2021} found that, in their sample of 75 galaxy clusters, 90\% of halos are hosted in disturbed objects, while only 10\% are in relaxed systems.

Cluster radio relics are usually found in the outskirts of merging galaxy clusters. They exhibit elongated morphologies and high degrees of polarisation above 1 GHz (up to 70\%, \citealt{Ensslin_1998, Bonafede_2014a, Loi_2019, deGasperin_2022}). The resolved spectral index in radio relics shows a gradient: it steepens towards the cluster centre and flattens towards the outskirts. Their size can reach up to $\sim$2 Mpc, and high-resolution observations have revealed filamentary structures within relics themselves \citep{diGennaro_2018, Rajpurohit_2020, deGasperin_2022, Rajpurohit_2022b, Rajpurohit_2022a}. The Largest Linear Sizes (LLS) and radio powers of relics are correlated, as well as the integrated spectral index and the radio power \citep{vanWeeren_2009b, Bonafede_2012, deGasperin_2014}.
Relics trace ICM shock waves with relatively low (M$<$3) Mach numbers \citep{Finoguenov_2010, Akamatsu_2013, Shimwell_2015, Botteon_2016}. The acceleration of electrons is believed to proceed via diffusive shock acceleration (DSA) in the ICM \citep{Ensslin_1998, Roettiger_1999}, in which particles scatter back and forth across the shock front gaining energy at every crossing. Nevertheless, this mechanism has been shown to be rather inefficient in accelerating electrons from the thermal pool (\citealt{Vazza_2014, Vazza_2016, Botteon_2020, Bruggen2020}; see \citealt{Brunetti_2014} for a review). Recently, it has been suggested that seed electrons could originate from the tails and lobes (driven by AGN outflows) of cluster radio galaxies \citep{Bonafede_2014a, vanWeeren_2017, Stuardi_2019}, which alleviates the requirements of high acceleration efficiencies at cluster shocks \citep[e.g.,][]{Markevitch_2005, Kang_2012, Botteon_2016, Eckert_2016, Kang_2017}.
In some cases, double relics have been detected on opposite sides of the cluster center
\citep[e.g.][]{Rottgering_1997, vanWeeren_2010, vanWeeren_2012, Bonafede_2012, deGasperin_2015}. In these clusters it is possible to constrain the merger history, providing important information about the formation processes of relics. 

Relativistic electrons with higher energies lose energy faster via synchrotron and Inverse Compton (IC) radiation. Therefore, the emission at high frequencies fades first. An old population of relativistic electrons, sometimes referred to as fossil plasma, can in some cases be "revived", e.g. by adiabatic compression \citep{Ensslin_2001}, leading to radio phoenices. These sources are characterized by steep and curved spectra ($\alpha <$ -1.5, \citealt{vanWeeren_2009, Clarke_2013, deGasperin_2015b, Mandal_2019}). They can exhibit different morphologies, even though in most cases they look elongated and filamentary \citep[e.g.][]{Slee_2001}. Compared to relics, phoenices are found at smaller distances from the cluster centre \citep{Feretti_2012} and they are smaller ($<$500 kpc). They can also be found in relaxed objects \citep[e.g.][]{vanWeeren_2011}, suggesting that major mergers are not strictly necessary for their formation. An alternative mechanism to re-accelerate old plasma was recently proposed by \citet{deGasperin_2017}. This was based on radio observations of Abell 1033 (A1033), in which long tails of radio bright plasma, generated by a radio galaxy moving within the cluster environment, are seen to brighten, in coincidence with a spectral index flattening. A possible explanation is that instabilities, generated by the interaction of the magnetically confined plasma in the tails and the turbulence in the surrounding medium, can lead to turbulent waves. These, in turn, are able to accelerate seed electrons through second-order Fermi mechanisms. This source was labeled Gently Re-Energised Tail (GReET) since the re-acceleration mechanism is barely efficient enough to balance the radiative losses of the electrons. Low-frequency observations of galaxy clusters are detecting an increasing number of tailed radio galaxies undergoing similar re-acceleration processes \citep[e.g.,][]{Cuciti_2018, Wilber_2018, Botteon_2021, Ignesti_2022, Brienza_2022, Pandge_2022}.

The contribution of the LOw-Frequency ARray (LOFAR, \citealt{vanHaarlem_2013}) is essential to achieve a better understanding of re-acceleration processes because of its unprecedented combination of sensitivity at low frequencies and resolution. Two surveys are currently being carried out with LOFAR. The LOFAR Two-Metre Sky Survey (LoTSS, \citealt{Shimwell_2017}, $<rms> \sim 100 \mu$Jy beam$^{-1}$) will observe the Northern Sky at a nominal frequency of 144 MHz (High Band Antennas, HBA) with a resolution of $\sim$6\arcsec. This survey is currently undergoing the second Data Release (DR2, \citealt{Shimwell_2022}), covering $\sim$6000 deg$^2$, while DR1 covered the HETDEX spring field\footnote{RA: 11h to 16h and Dec: 45$\degree$ to 62$\degree$ \citep{Hill_2008}.} \citep{Shimwell_2019}. Similarly to LoTSS, the Lofar LBA Sky Survey (LoLSS, \citealt{deGasperin_2021}) will observe the Northern Sky at a nominal frequency of 54 MHz (Low Band Antennas, LBA) with a resolution of $\sim$15\arcsec, with the first Data Release covering HETDEX \citep{deGasperin_2021}. The data reduction and calibration of HETDEX is now completed. The analysis of all known galaxy clusters in this field at 54 MHz, similarly to what was done at 144 MHz by \citealt{vanWeeren_2021} (hereafter VW21), will be carried out in a forthcoming publication (Pasini et al. in prep.). 

In this paper, we present LOFAR, Upgraded Giant Metrewave Radio Telescope (uGMRT) and Very Large Array (VLA) observations of one of the most interesting HETDEX clusters, Abell 1550 (alternatively PSZ2G133.60+69.04, hereafter A1550). This is a dynamically disturbed cluster located at $z \sim 0.254$, with $M_{500}$\footnote{Mass within R$_{500}$, defined as the radius at which the medium density is 500 times the critical density of the Universe.} $\sim 5.88 \times 10^{14} M_{\odot}$ \citep{Planck_2016}. According to \citet{Wen_2013}, the cluster hosts about a hundred confirmed member galaxies within $R_{200} \sim 2$ Mpc. Extended emission was already claimed by \citet{Govoni_2012} through 1.4 GHz Very Large Array (VLA) data in D and C configuration. LOFAR observations at 144 MHz confirm the presence of diffuse emission (VW21, \citealt{Botteon_2022}), albeit on a greater scale of 1.8 Mpc. The difference in size is due to the high sensitivity and low frequency coverage of LOFAR. A \textit{Chandra} 7 ks archival observation is also available, while the ROSAT All-Sky Survey (RASS) measures an X-ray luminosity of $L_X \sim 3.5 \times 10^{44}$ erg s$^{-1}$ in the 0.1-2.4 keV band \citep{Bohringer_2000}. 
In Sec.~\ref{sec:analysis} we discuss the data reduction and calibration, while in Sec.~\ref{sec:results} we present our results which are then discussed in Sec.~\ref{sec:discussion}. Finally, in Sec.~\ref{sec:conclusions} we summarise our main conclusions. We assume a standard $\Lambda$CDM cosmology with H$_0 = 70$ km s$^{-1}$ Mpc$^{-1}$, $\Omega_\Lambda = 0.7$ and $\Omega_{\text{M}} =  1-\Omega_\Lambda  = 0.3$, and errors are at 68$\%$ confidence level (1$\sigma$).

%%%%%%%%%%%%%%%%%%%%%%%%%%%%%%%%%%%%%%%%%%%%%%%%%%%%%%%%%%%%%%%%%%%%%%%%%%%%%%

\section{Data analysis}
\label{sec:analysis}

\begin{figure*}
    \includegraphics[scale=1]{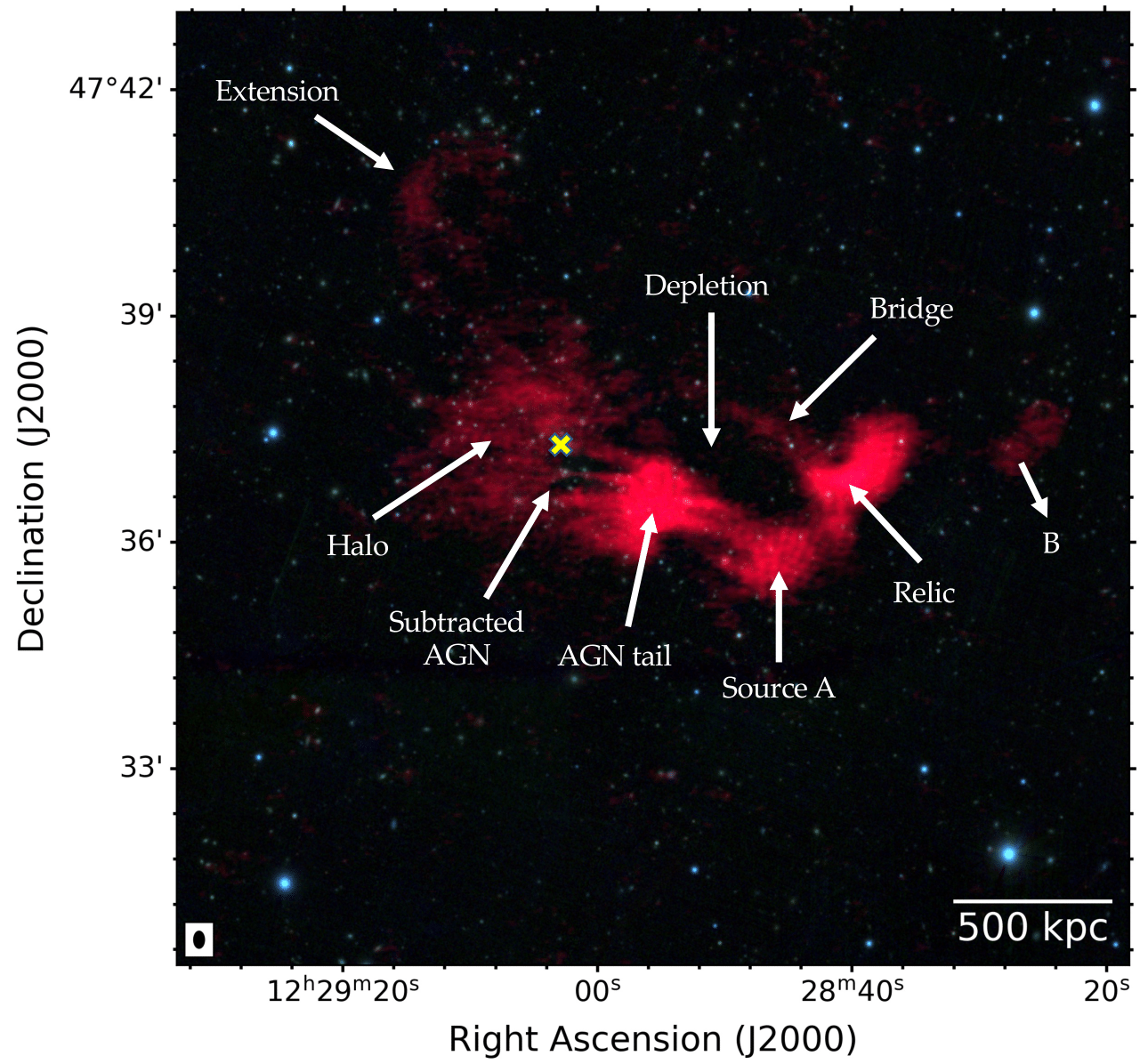}
	\caption{A1550 seen in 144 MHz emission (red) as detected by LOFAR overlaid on the PanSTARSS optical (RGB filters) image. The radio image has a resolution of 13\arcsec$\times$8\arcsec, and the beam is shown on the bottom right. The most prominent structures that are discussed in the text are labeled. Compact sources are subtracted to enhance diffuse emission as described in Sec.~\ref{sec:LBAcalib}. The yellow cross marks the cluster X-ray centre, defined as the peak of the emission and estimated from the \textit{Chandra} image presented in Sec. \ref{sec:chandra}.}
	\label{fig:overlay}
\end{figure*}

\subsection{LBA observations}
\label{sec:LBAcalib}

The galaxy cluster A1550 was observed by LOFAR in the 42 -- 66 MHz band as part of the HETDEX field, which is the first region of the sky targeted by LoLSS \citep{deGasperin_2021}. The cluster is covered by four different pointings (see Table \ref{tab:pointings} for details), for a total of 27 hours after data reduction and calibration.

Data were calibrated using the automated Pipeline for LOFAR LBA (PiLL\footnote{Publicly available at https: //github.com/revoltek/LiLF}). The pipeline was independently run for each of the pointings and it is described in detail in \citet{deGasperin_2021}. Here, we summarise the main steps. First, phase and bandpass solutions of the calibrator are determined and transferred to the target field. Then, direction-independent calibration of the target field is performed to correct for the direction-averaged ionospheric delays, Faraday rotation and second-order beam errors. At this step, wide-field images are produced and direction-dependent (DD) errors, mainly due to ionospheric inhomogeneities, still affect the data. To correct these errors, we adopt the following strategy for direction-dependent calibration:  we subtract all sources from the visibilities using the model obtained during direction-independent calibration. Then, we re-add the brightest source to the data and derive calibration solutions for the direction corresponding to this source  using self-calibration. Using these solutions and the improved source model, we subtract the source again, this time more accurately. This cycle is repeated for all sufficiently bright sources (DD sources) in the field of view (FoV). We then divide the FoV into facets whose position and shape depend on the position of DD sources. Each facet is calibrated for DD effects during imaging using the correction of the respective calibrator. This is done using {\ttfamily DDFacet} \citep{Tasse_2018}, and produces a DD-calibrated wide-field image. To further improve image quality, we repeat the steps of the direction-dependent calibration, starting from the DD-calibrated image.

It is still possible to optimize the calibration for a specific target of interest (not included among DD sources), as in our case for A1550. Here, we adapted the extraction process described in VW21 to LBA data. The procedure has already been tested in \citet{Biava_2021}. During DD calibration, it is assumed that DD effects are uniform across the facet. By selecting a smaller region (typically 15$'$ -- 20$'$) around the target, it is possible to relax this assumption and improve calibration (see also VW21 for more details). The extraction region should contain enough flux density to allow a good calibration of the target.

The procedure consists of the following steps: first, we subtract all sources outside the extraction region. We then shift the phase centre of the observation to the centre of the region, and we average the data in frequency and time. We apply a correction to account for the LOFAR primary beam response at the new phase centre through IDG (Image-Domain Gridder, \citealt{vanderTol_2019}). Pointings for which the beam response drops below 30\% are excluded, while all the others are combined and weighted so that pointings with higher beam sensitivity contribute more to the extracted dataset. Then we perform self-calibration cycles as in VW21 to further reduce the noise and improve the quality of the images.
Final imaging is carried out using {\ttfamily WSClean} \citep{Offringa_2014} applying suitable weighting and tapering of the visibilities in order to obtain images at different resolutions. Also we use multi-scale deconvolution \citep{Offringa_2016}, and we assume that flux errors are $\sim$10\%, as for LoLSS \citep{deGasperin_2021}.

In some cases we needed to highlight the emission from diffuse sources. When specified, we subtracted compact sources as follows: first, we produced a high-resolution image by applying {\ttfamily Briggs -0.6} and cutting visibilities below 100$\lambda$, which correspond to angular scales above 35$'$. We chose the resolution such that only compact sources with a Largest Linear Size (LLS) below a given threshold are imaged. Through the {\ttfamily predict} option of WSClean, the clean components of the previous image are stored as model. Finally, we subtract this model out of the \textit{uv}-data. This leaves us with only the visibilities of the diffuse emission.

\begin{table*}
	\centering
	\footnotesize
	\begin{tabular}{c c c c c c c}
		\hline
		\hline
		Observation & Central frequency & RA & DEC & Total exposure & Target distance & Antenna configuration\\
		& & [\textit{hh:mm:ss}] & [\textit{deg:mm:ss}] & & [deg] &\\
		\hline
		P183+47 & 54 MHz & 12:13:09.65 & +47:15:17.30 & 6hr & 2.75 & LOFAR LBA\_OUTER\\
		P186+50 & 54 MHz & 12:27:40.70 & +49.46.59.83 & 7hr & 2.09 & LOFAR LBA\_OUTER\\
		P187+47 & 54 MHz & 12:28:25.34 & +47.16.29.19 & 7hr & 0.45 & LOFAR LBA\_OUTER\\
		P190+47 & 54 MHz & 12:43:41.37 & +47.17.41.33 & 7hr & 2.48 & LOFAR LBA\_OUTER\\
		P23Hetdex & 144 MHz & 12:21:08.60 & +47:29:24.00 & 8hr & 1.30 & LOFAR HBA\_DUAL\_INNER\\
		P26Hetdex & 144 MHz & 12:29:37.60 & +49:44:24.00 & 8hr & 2.13 & LOFAR HBA\_DUAL\_INNER\\
		P27Hetdex & 144 MHz & 12:38:06.70 & +47:29:24.00 & 8hr & 1.55 & LOFAR HBA\_DUAL\_INNER\\
		12435 & 400 MHz & 12:28:54.00 & +47:36:44.00 & 3hr & 0.03 & uGMRT BAND 3\\
		18A-172$^{*}$ & 1.4 GHz & 12:29:12.7 & +47:42:25.38 & 1hr & 0.09 & JVLA C array\\
		18A-172$^{*}$ & 1.4 GHz & 12:29:12.7 & +47:42:25.38 & 1hr & 0.09 & JVLA D array\\
		\hline
	\end{tabular}
	\caption{The table shows the details of the LBA, HBA, uGMRT and JVLA observations used for the analysis of A1550. From left to right: pointing/observation name, nominal frequency of the observation, coordinates of the phase centre, distance of A1550 from the phase centre and configuration of the antennae of the related observation. $^{*}$: \textit{Project name}.} \label{tab:pointings}
\end{table*}

\begin{figure*}
	\centering
    \includegraphics[scale=0.835]{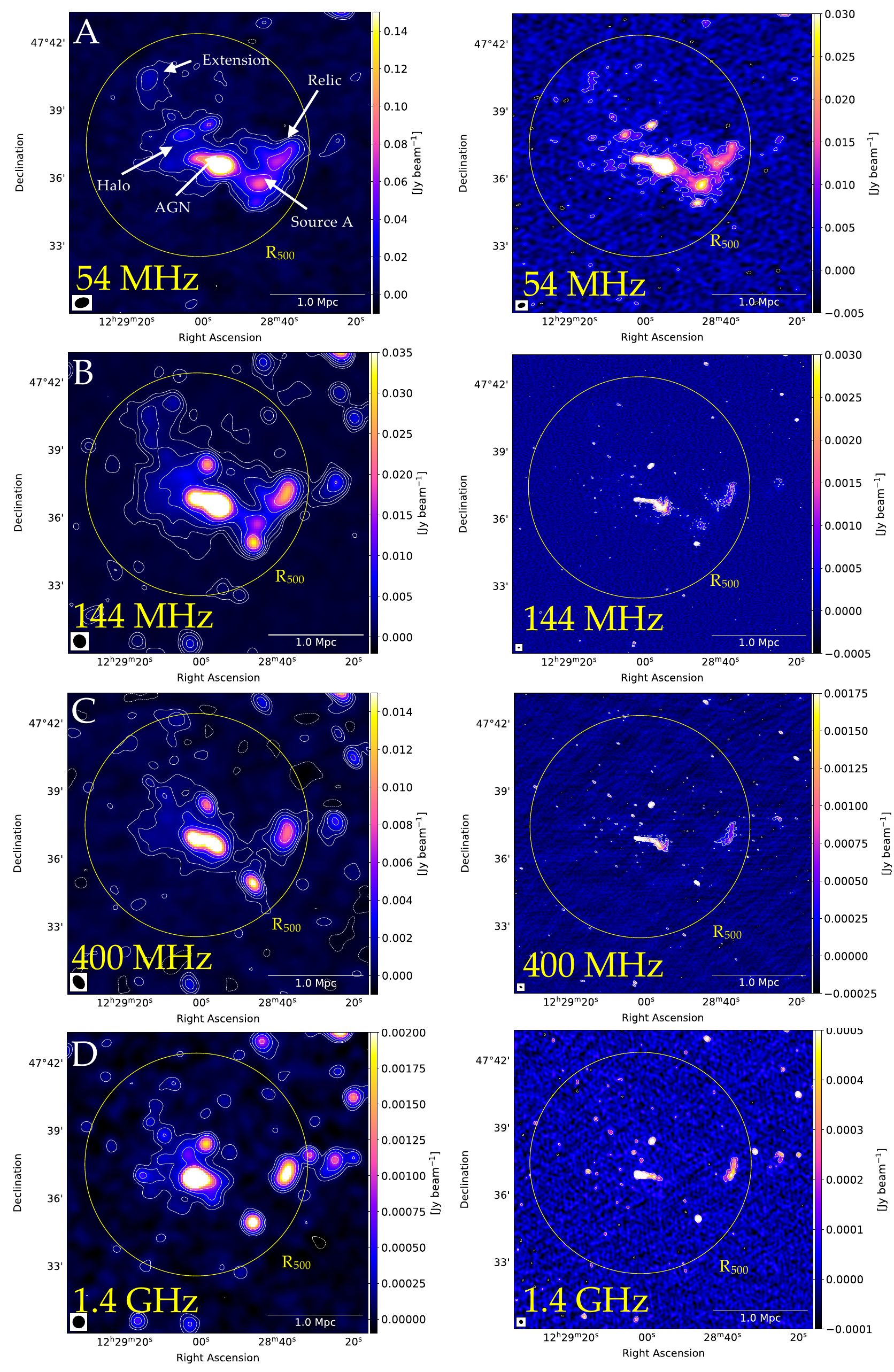}
	\caption{First to last row: 54, 144, 400 MHz and 1.4 GHz, low- (left column) and high-resolution (right column) images of A1550, produced by tapering visibilities at 30\arcsec and applying {\ttfamily Briggs -0.6}, respectively. The yellow circle denotes $R_{500} = 1173$ kpc \citep{Botteon_2022}. \textit{Row A, left panel}: 54 MHz low-resolution image. The beam is 38\arcsec$\times$25\arcsec, with \textit{rms} noise $\sigma \sim$1.6 mJy beam$^{-1}$. Contours are at [-3, 3, 6, 12, 24] $\times \sigma$. \textit{Row A, right panel}: 54 MHz high-resolution image. The beam is 18\arcsec$\times$11\arcsec, with \textit{rms} noise $\sigma \sim$1.5 mJy beam$^{-1}$. \textit{Row B, left panel}: 144 MHz low-resolution image. The beam is 35\arcsec$\times$32\arcsec, with \textit{rms} noise $\sigma \sim$0.14 mJy beam$^{-1}$. \textit{Row B, right panel}: 144 MHz high-resolution image. The beam is 8\arcsec$\times$4\arcsec, with \textit{rms} noise $\sigma \sim$69$ \ \mu$Jy beam$^{-1}$. \textit{Row C, left panel}: 400 MHz low-resolution image. The beam is 38\arcsec$\times$26\arcsec, with \textit{rms} noise $\sigma \sim$0.28 mJy beam$^{-1}$. \textit{Row C, right panel}: 400 MHz high-resolution image. The beam is 6\arcsec$\times$3\arcsec, with \textit{rms} noise $\sigma \sim 60 \ \mu$Jy beam$^{-1}$. \textit{Row D, left panel}: 1.4 GHz low-resolution image. The beam is 32\arcsec$\times$32\arcsec, with \textit{rms} noise $\sigma \sim 26 \ \mu$Jy beam$^{-1}$. \textit{Row D, right panel}: 1.4 GHz high-resolution image. The beam is 13\arcsec$\times$12\arcsec, with \textit{rms} noise $\sigma \sim 18 \ \mu$Jy beam$^{-1}$.
	}
	\label{fig:br0LOFAR}
\end{figure*}

\subsection{HBA observations}
\label{sec:HBAcalib}

A1550 was observed by LOFAR in the frequency range 120-166 MHz as part of the second Data Release (DR2) of LoTSS \citep{Shimwell_2022}. The cluster is covered by three pointings (see Table \ref{tab:pointings} for details), for a total of $\sim$24 hours. The data were processed with a set of fully automated pipelines developed by the LOFAR Surveys Key Science Project team: {\ttfamily prefactor} \citep{vanWeeren_2016, Williams_2016, deGasperin_2019} and {\ttfamily ddf-pipeline} \citep{Tasse_2021}. These pipelines are able to correct for both direction-independent and direction-dependent effects. The pipelines also include flagging of the radio frequency interference (RFI). Then the \textit{uv}-data is averaged in time and frequency, and complex gains, clock offsets and phase delays are obtained and applied to each station. The calibration of direction-dependent effects (DDE) is then performed using the same facet method discussed for LBA.

In this work, we exploit the same datasets which were recently presented in \citet{Botteon_2022}. New images were produced through {\ttfamily WSClean} with different sets of parameters, depending on our purposes, applying multiple weighting schemes and visibility tapering. The flux density scale was aligned with the LoTSS-DR2 data release \citep{Botteon_2022}, where the flux calibration uncertainty is estimated to be $\sim$10\% \citep{Hardcastle_2021}. Where explicited, compact-source subtraction was performed exploiting the same method described in Sec. \ref{sec:LBAcalib}.

\subsection{GMRT observation}
\label{sec:GMRTcalib}

A1550 was observed by uGMRT on September 17th, 2020, for a total integration time of 3 hours (ObsID = 12435, PI V. Cuciti). Observations were carried out in band 3 (330-500 MHz), with nominal frequency 400 MHz. 3C286 was used as absolute flux density calibrators. Data reduction and calibration were carried out using the Source Peeling and Atmospheric Modeling ({\ttfamily SPAM}) pipeline \citep{Intema_2009, Intema_2017}, which corrects for ionospheric effects and removes direction-dependent gain errors. Direction-dependent gains were derived through bright sources in the field of view. Finally, data were corrected for the system temperature variations between calibrators and target. Imaging is carried out through {\ttfamily WSClean} applying different weightings and visibility tapering. The flux uncertainty is assumed to be 6\% \citep{Chandra_2004}. When specified, we applied the same compact-source subtraction procedure described in Sec. \ref{sec:LBAcalib}.

\subsection{JVLA observations}
\label{sec:vla}

We obtained JVLA observations of A1550 in the 1-2 GHz JVLA L-band, in C and D antenna configuration (Project ID = 18A-172, PI R. J. van Weeren). The former was performed on January 19, 2019 for a total of 1 hour, pointed at RA=12h29m12.75s DEC=47$\degree$42$'$25.4\arcsec, while the latter is a 1 hour observation performed on 20 September, 2018 and pointed at the same coordinates. Data reduction was carried out with the National Radio Astronomy Observatory (NRAO) Common Astronomy Software Applications package (CASA, version 6.1.2.7), exploiting 3C286 as primary calibrator for both observations, while J1219+4829 was used as phase calibrator. First, RFI and bad visibilities for calibrators and target were flagged. Amplitude and phase solutions were derived from the calibrators, and applied to the target. For the purpose of self-calibration, phase-solutions are then calculated from the model of the target, produced by a first, shallow imaging, and computed on a given timescale for which the phase is assumed to be constant. After applying these solutions, the cycle is repeated in order to increase the Signal-to-Noise (S/N) ratio and improve the quality of the image, decreasing the interval at each cycle. After this step, datasets were accurately combined to achieve longer integration time and increase the sampling of the \textit{uv}-plane. Final imaging is carried out with {\ttfamily WSClean}, and the flux uncertainty is assumed to be 5\%. When specified, we applied the same compact-source subtraction procedure described in Sec. \ref{sec:LBAcalib}. Finally, for polarisation calibration, the leakage response was determined using the unpolarised calibrator 3C147. The absolute position angle (the R-L phase difference) was corrected using the polarised calibrator 3C286. The polarisation intensity ($P$) and polarisation angle $(\Psi$) maps were derived from Stokes Q and U maps:

\begin{equation}
\centering
\begin{split}
P= \sqrt{Q^{2}+U{^2}}, \\
\Psi=\frac{1}{{2}}\tan ^{ - 1} {\frac{U}{Q}}
\end{split}
\end{equation}

Finally, the fractional polarisation map was obtained as:

\begin{equation}
\centering
p= \frac{P}{I},
\end{equation}

where $I$ and $P$ are the total intensity and polarization intensity, respectively, of the source.

\subsection{\textit{Chandra} observation}
\label{sec:chandra}

The \textit{Chandra} data of A1550 (obsID 11766) were taken with the Advanced CCD Imaging Spectrometer (ACIS) in S configuration, operating in VFAINT mode, with a short exposure time of $\sim$7 ks. The ACIS-S configuration consists of an array of 6 chips. We use the ACIS-S3 chip, where the aimpoint of the telescope lies. Unfortunately, the cluster emission covers a projected region larger than the CCD area (8.3$' \times 8.3'$). For this reason, part of the cluster emission is out of the FoV and cannot be studied. The maximal area that can be reached lies within a radius of 4 arcmin from the cluster centre, which corresponds to $\sim$0.8R$_{500}$.

The data were reprocessed with CIAO 4.11\footnote{https://cxc.harvard.edu/ciao/} using CALDB 4.8.4.1. The {\ttfamily Chandra\_repro} script was executed to perform the standard calibration process. Background flares were removed and the {\ttfamily Blanksky} template files, filtered and normalized to the count rate of the source in the hard X-ray band (9-12 keV), were exploited to model the background contribution to the emission. Finally, point sources were identified and removed using the CIAO task {\ttfamily WAVDETECT} with a default significance threshold of 10$^{-6}$. The final exposure time is 6948 s. Exposure-corrected images were produced in the 0.5-7 keV band.

%%%%%%%%%%%%%%%%%%%%%%%%%%%%%%%%%%%%%%%%%%%%%%%%%%%%%%%%%%%%%%%%%%%%%%%%%%%%%%

\section{Results}
\label{sec:results}

\subsection{HBA and LBA comparison}

In Fig.~\ref{fig:overlay} we show the radio emission as detected at 144 MHz by HBA, overlaid on the Panoramic Survey Telescope and Rapid Response System (PanSTARRS, \citealt{Chambers_2016}) image of A1550. Projected at the cluster centre, an head-tail (HT) radio galaxy (labelled as AGN+tail in the figure) dominates the emission. The galaxy is embedded within the giant radio halo originally detected by \citet{Govoni_2012}. In the South-West, a roundish patch of emission (A) is observed, without a clear optical counterpart. North-West of source A, we clearly detect a source which resembles a radio relic, as already hinted at in VW21 and \citet{Botteon_2022}. From this image, it is not clear whether the putative relic and source A are part of the same structure. From here, two filaments extend to the East and West directions. The eastern filament (bridge) looks connected to the halo, as already found at 144 MHz by VW21 from low-resolution images. The western filament, labeled B, has no clear counterpart as well, even though its morphology suggests that it could be a radio galaxy, possibly unrelated to the diffuse emission. An elliptical depletion of radio emission is observed between the bridge, the head-tail and source A. Finally, in the North-East region of the cluster, we observe an arc-shaped extension of emission. As already discussed in VW21, it lies in front (at least in projection) of a group of galaxies with the same redshift of A1550. Low-resolution images at 144 MHz showed that it is likely connected to the halo (VW21 and \citealt{Botteon_2022}). We will return on this throughout this paper.

\begin{figure*}
	\centering
	\includegraphics[scale=0.95]{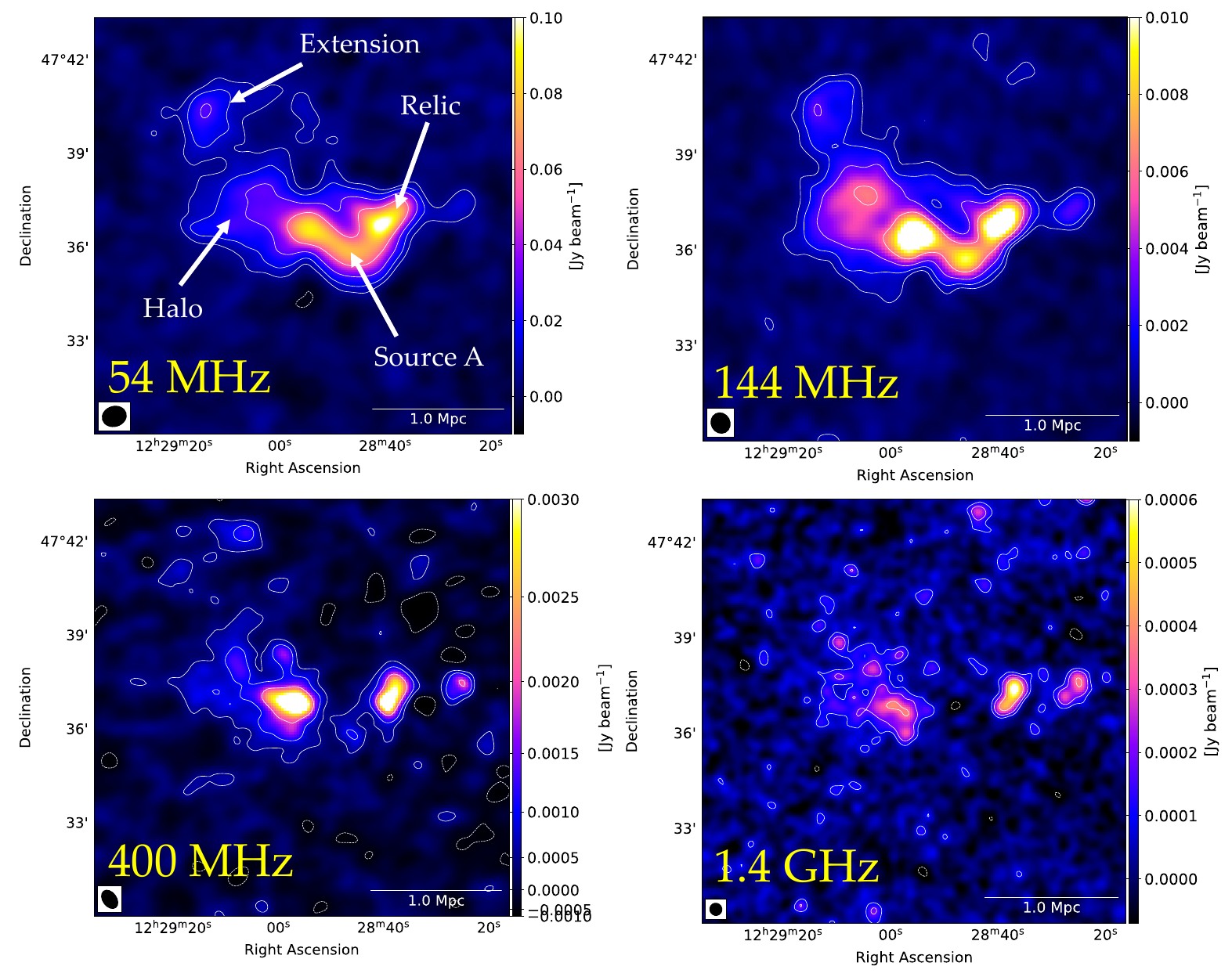}
	\caption{\textit{Top left}: 54 MHz source-subtracted image of A1550 obtained applying {\ttfamily Briggs -0.3} and tapering the visibilities to 30\arcsec. The beam, shown on the bottom left, is 46\arcsec$\times$36\arcsec, with \textit{rms} noise $\sigma \sim$ 1.9 mJy beam$^{-1}$. Contours are at -3, 3, 6, 12, 24 $\times \sigma$. \textit{Top right}: 144 MHz source-subtracted image of A1550 obtained applying {\ttfamily Briggs -0.3} and tapering the visibilities to 30\arcsec. The beam is 36\arcsec$\times$35\arcsec, with \textit{rms} noise $\sigma \sim$ 0.14 mJy beam$^{-1}$. Contours are the same as above. \textit{Bottom left}: 400 MHz source-subtracted image of A1550 obtained applying {\ttfamily Briggs -0.3} and tapering the visibilities to to 30\arcsec. The beam is 36\arcsec$\times$26\arcsec, with \textit{rms} noise $\sigma \sim$ 0.23 mJy beam$^{-1}$. \textit{Bottom right}: 1.4 GHz source-subtracted image of A1550 obtained applying {\ttfamily Briggs -0.3} and tapering the visibilities to to 30\arcsec. The beam is 32\arcsec$\times$32\arcsec, with \textit{rms} noise $\sigma \sim$ 23 $\mu$Jy beam$^{-1}$.}
	\label{fig:subtractedLOFAR}
\end{figure*}

With the purpose of unveiling the nature and morphology of each of these structures, we produced images at different frequencies and resolution. In the left panels of the first two rows in Fig.~\ref{fig:br0LOFAR}, we show the cluster as detected at 54 MHz and 144 MHz by tapering visibilities at 30\arcsec. The head-tail dominates the radio emission, with integrated flux density inside 3$\sigma$ contours\footnote{If not specified otherwise, all flux densities are calculated within 3$\sigma$ contours.} of $S_{144 \rm{MHz}} = 264 \pm 26$ mJy and $S_{54 \rm{MHz}} = 707 \pm 71$ mJy, respectively. The putative relic and halo are clearly visible at both frequencies, and the morphology and orientation of the NE extension suggest that it is connected to the halo. The bridge connecting the candidate relic to the halo, while resolved in HBA, is hardly distinguishable in the LBA image due to lower resolution. 

In order to resolve all the detected structures, we produced high-resolution images by setting {\ttfamily Briggs -0.6} and excluding baselines shorter than 100$\lambda$. The halo almost disappears at both frequencies (see right panels of Fig.~\ref{fig:br0LOFAR}), while we are still able to see the candidate relic and source A. At 144 MHz, the emission that connects source A to the AGN/halo is split into two arms, which will be better defined and analysed in Sec. \ref{sec:gentle}. The lack of optical counterparts, together with the shape and orientation of the head-tail radio galaxy and relic, might suggest that source A could be constituted by re-accelerated electrons (see Sec. \ref{sec:gentle}).

We then subtracted compact sources as described in Sec.~\ref{sec:LBAcalib} on the HBA and LBA observations and tapered visibilities at 30\arcsec, with the purpose of imaging more accurately the diffuse emission. The resulting images are shown in Fig.~\ref{fig:subtractedLOFAR}. Without the AGN, it is much easier to assess the different structures observed in A1550, and we can therefore estimate their physical properties, such as flux density and extent, more accurately.

To estimate the flux density of the halo, we used the Halo Flux Density CAlculator (Halo-FDCA, \citealt{Boxelaaar_2021}) on the subtracted maps. This code fits the surface brightness of radio halos to 2D exponential models using Bayesian inference, and calculates the flux density analytically. The plots of the halo models and masks used for the fit are shown in Appendix \ref{fdca_appendix}. After accounting for the contribution of the radio galaxy tail and masking both source A and NE extension, we measure a flux density for the halo of $S_{144 \rm {MHz}} = 108 \pm 11$ mJy, which is in agreement with what we measure from 2$\sigma$ contours, and a maximum extent of $\sim$1 Mpc. The fit also provides an estimate of the \textit{e}-folding radius of $r_e = 183 \pm 4$ kpc. We note that VW21 reported a flux density of $S_{144 \rm {MHz}} = 129 \pm 26$ mJy, while \citet{Botteon_2022} provides an estimate of $S_{144 \rm {MHz}} = 145 \pm 18$ mJy. The difference with the previous results is likely related to the fact that although VW21 and \citet{Botteon_2022} also used FDCA, they performed the fit with an elliptical model, while we used a circle model. In \citet{Botteon_2022}, this resulted in radii of $377 \pm 5$ and $201 \pm 4$ kpc, for the major and minor axes respectively. Given the wealth of radio data studied in this work, the circle model appears to better suit the morphology of the halo emission. From the LBA image, we find $S_{54 \rm {MHz}} = 498 \pm 57$ mJy and projected LLS$\sim$1.2 Mpc. This implies a relatively steep spectral index of $\alpha_{54 \rm MHz}^{144 \rm MHz} = -1.6 \pm 0.2$. The \textit{e}-folding radius is estimated to be $r_e = 177 \pm 6$ kpc, consistent within errors with HBA.

\begin{figure}
	\centering
	\includegraphics[scale=0.5]{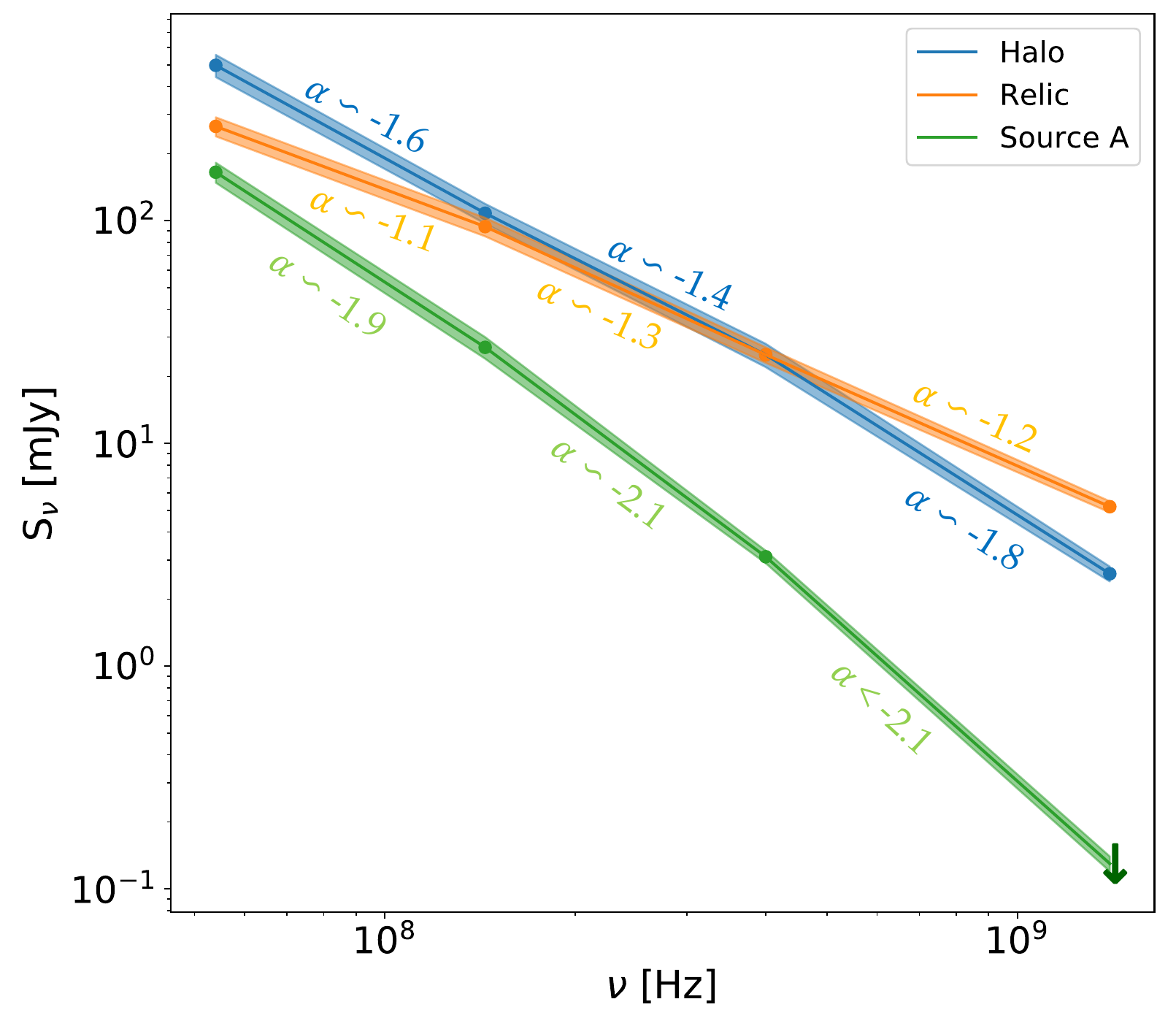}
	\caption{Flux density as a function of frequency for the halo, the relic and source A at the four frequencies covered in our analysis. Colored areas denote flux density errors. Note that the flux density at 1.4 GHz for source A is a 3$\sigma$ upper limit, with $\sigma$ being the \textit{rms} noise of the 1.4 GHz image.}
	\label{fig:spectra}
\end{figure}

The putative relic shows $S_{144 \rm {MHz}} = 94 \pm 9$ mJy, with a projected length of $\sim$500 kpc. With LBA, we measure $S_{54 \rm {MHz}} = 265 \pm 26$ mJy. This implies $\alpha_{54 \rm MHz}^{144 \rm MHz} = -1.1 \pm 0.2$, which is typical of relics (VW21). Finally, for source A we find $S_{144 \rm {MHz}} = 27 \pm 3$ mJy and $S_{54 \rm {MHz}} = 165 \pm 17$ mJy, resulting in a steep spectrum with $\alpha_{54 \rm MHz}^{144 \rm MHz} = -1.9 \pm 0.2$. A more detailed investigation of spectral indices in this cluster is presented in Sec.~\ref{sec:spindex}. In Table \ref{tab:flux}, we summarise the flux densities at different frequencies of the halo, candidate relic and source A. Their spectra are plotted in Fig.~\ref{fig:spectra}.

\begin{table*}
	\centering 
	\begin{tabular}{c c c c c c c c}
		\hline
		\hline
		Source & S$_{54 \rm MHz}$ & S$_{144 \rm MHz}$ & S$_{400 \rm MHz}$ & S$_{1.4 \rm GHz}$ & $\alpha_{54 \rm MHz}^{144 \rm MHz}$ & $\alpha_{144 \rm MHz}^{400 \rm MHz}$ & $\alpha_{400 \rm MHz}^{1.4 \rm GHz}$ \\
		& [mJy] & [mJy] & [mJy] & [mJy] & & & \\
		\hline
		Halo & 498 $\pm 57$ & 108 $\pm$ 11 & 25 $\pm$ 3 & 2.6 $\pm$ 0.2 & $-$1.6 $\pm$ 0.2 & $-$1.4 $\pm$ 0.1 & $-$1.8$\pm$ 0.1 \\
		Relic & 265 $\pm$ 26 & 94 $\pm$ 9 & 25 $\pm$ 2 & 5.2 $\pm$ 0.3 & $-$1.1 $\pm$ 0.2 & $-$1.3 $\pm$ 0.2 & $-$1.2 $\pm$ 0.2 \\
		Source A & 165 $\pm$ 17 & 27 $\pm$ 3 & 3.1 $\pm$ 0.2 & $<0.13^*$ & $-$1.9 $\pm$ 0.2 & $-$2.1 $\pm$ 0.1 & / \\
		\hline
	\end{tabular}
	\caption{Flux densities and integrated spectral indices of halo, relic and source A at different radio frequencies.$^*$: 3$\sigma$ upper limit, with $\sigma$ being the \textit{rms} noise of the 1.4 GHz image.} 
	\label{tab:flux}
\end{table*}

\subsection{The cluster at 400 MHz}

In the third row of Fig.~\ref{fig:br0LOFAR} we show the 400 MHz low- and high-resolution images of A1550. The head-tail radio galaxy dominates the emission. To the West of the head-tail, the putative relic is clearly visible, with a flux density $S_{\rm 400 MHz} = 25 \pm 2$ mJy. Hence, its integrated spectral index between 144 and 400 MHz is $\alpha_{144\rm MHz}^{400\rm MHz} = -1.3 \pm 0.2$, consistent with what estimated at LOFAR frequencies. At 400 MHz we do not see the bridge, suggesting that it has a steep spectrum. Furthermore, it is not trivial to understand whether the emission between the head-tail and the candidate relic belongs to source A or comes from these two sources. Instead, around the radio galaxy we clearly observe the radio halo.

As before, we then subtracted compact sources and tapered visibilities to a resolution of 30\arcsec, with the result shown in the bottom-left panel of Fig.~\ref{fig:subtractedLOFAR}. We observe what is probably a hint of emission of source A, even though it is detected barely at 3$\sigma$ significance. At this frequency, we measure a maximal extent of the halo of $\sim$950 kpc and a flux density (with Halo-FDCA) of $S_{\rm 400 MHz} = 25 \pm 3$ mJy. This translates into a spectral index between 144 MHz and 400 MHz of $\alpha_{144\rm MHz}^{400\rm MHz} = -1.4\pm 0.2$, consistent within errors with the one estimated between 54 and 144 MHz. The \textit{e}-folding radius is $r_e = 171 \pm 10$ kpc, which is comparable to those estimated at lower frequencies.

Finally, the flux density of source A at this frequency is $S_{\rm 400 MHz} = 3.1 \pm 0.2$ mJy. The integrated spectral index between 144 and 400 MHz is as steep as $\alpha_{144\rm MHz}^{400\rm MHz} = -2.1 \pm 0.2$. We therefore observe a steepening at higher frequencies, hinting that the spectrum of source A might be significantly curved.

\subsection{The cluster at 1.4 GHz}

\begin{figure*}
	\centering
    \includegraphics[scale=0.6]{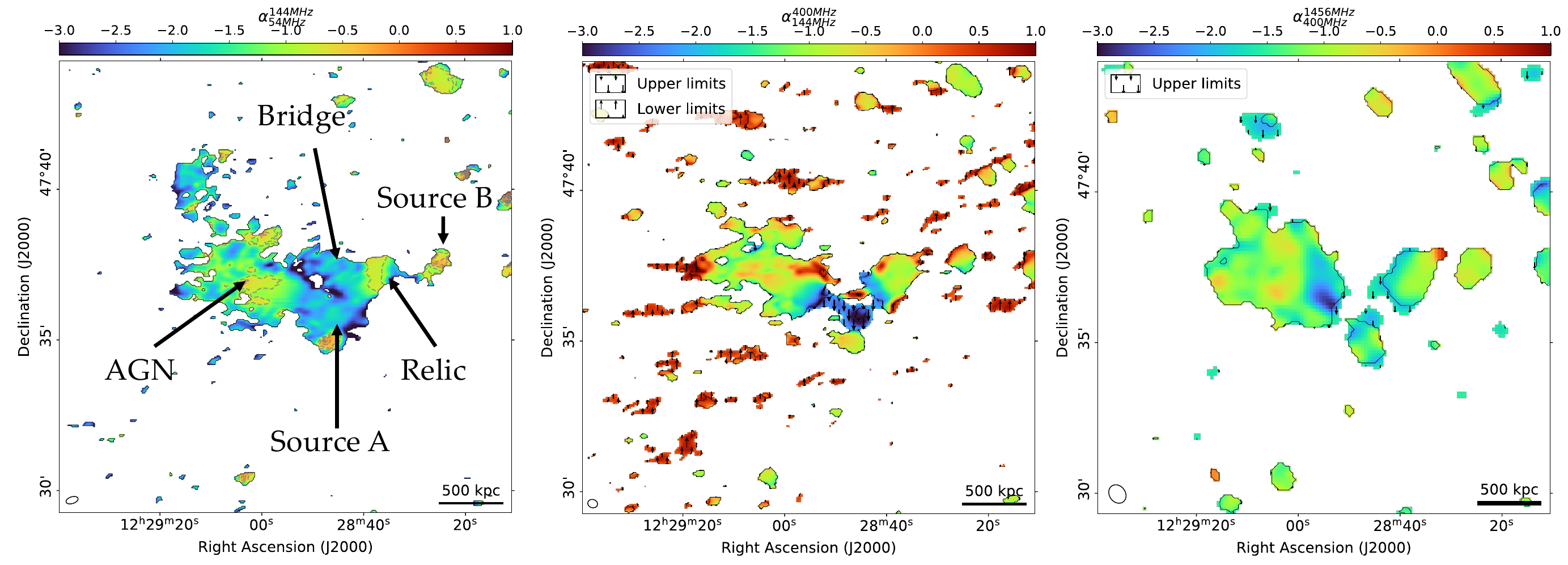}
    \includegraphics[scale=0.605]{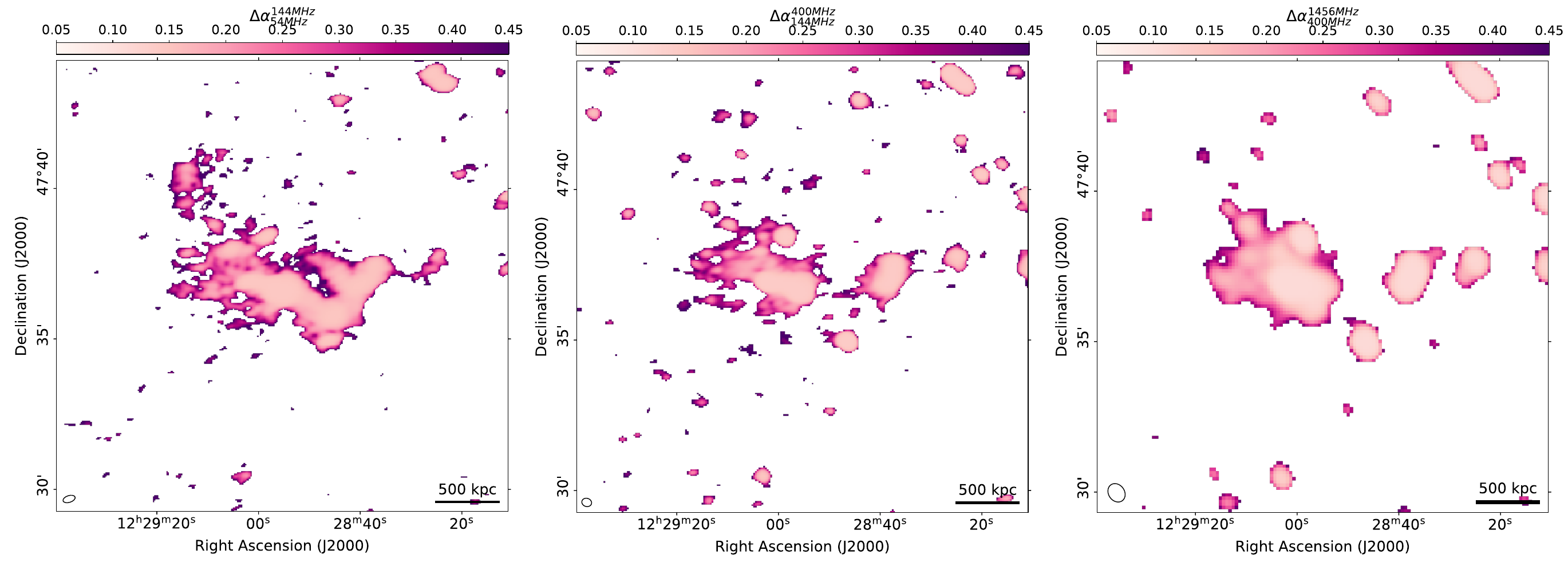}
	\caption{\textit{Top Left}: Spectral index map between 54 MHz and 144 MHz, generated by combining total intensity maps produced with matching \textit{uv}-cut 80$\lambda$-14k$\lambda$. Contours are at 3$\sigma$ from the LBA map. The beam is 25\arcsec$\times$13\arcsec. \textit{Top Middle}: Spectral index map between 144 MHz and 400 MHz, generated by combining total intensity maps produced with matching \textit{uv}-cut 80$\lambda$-14k$\lambda$. The beam is 20\arcsec$\times$17\arcsec. \textit{Top Right}: Spectral index map between 400 MHz and 1.4 GHz, generated by combining total intensity maps produced with matching \textit{uv}-cut 140$\lambda$-14k$\lambda$. The beam is 38\arcsec$\times$32\arcsec. \textit{Bottom Left}: Spectral index error map between 54 MHz and 144 GHz. \textit{Bottom Middle}: Spectral index error map between 144 MHz and 300 MHz. \textit{Bottom Right}: Spectral index error map between 400 MHz and 1.4 GHz.}
	\label{fig:spindexLOFAR}
\end{figure*}

In the bottom panels of Fig.~\ref{fig:br0LOFAR} we show A1550 at 1.4 GHz as observed by JVLA, at low and high resolution. At this frequency the halo appears less extended, as expected, but still clearly visible. We also detect the candidate relic, while we observe no emission from source A and the bridge which are detected with LOFAR. North-East to the halo, there is a hint of what probably is the high-frequency counterpart of the extension. As above, we performed the subtraction of compact sources, with the caveat that JVLA array D provides, by definition, low-resolution, making the subtraction trickier. The result is shown in the bottom right panel of Fig.~\ref{fig:subtractedLOFAR}.

The total extent of the halo is $\sim$900 kpc. Similarly to what observed at 144 MHz, the inconsistency with the previous measurement ($\sim$1.4 Mpc) by \citet{Govoni_2012}, performed exploiting older VLA datasets, might be due to the exclusion of the NE extension (allowed by the higher resolution), other than to a better \textit{uv}-coverage of the most recent observations. We estimate a total flux density for the halo (with Halo-FDCA\footnote{Note that, due to a less trivial source subtraction for JVLA, we had to use a larger number of masks for the flux density estimate. See also Appendix.}) of $S_{\rm 1.4GHz} = 2.6 \pm 0.2$ mJy, implying $\alpha_{\rm 400MHz}^{\rm 1.4GHz} = -1.8 \pm 0.1$, consistent within errors with the spectral index determined at LOFAR frequencies. \citet{Govoni_2012} provided an estimate of $\sim7.7$ mJy at the same frequency. However, their calculation included, both, the NE extension and source A. Even after source subtraction we do not detect the latter, confirming that it might have a steep spectrum. The \textit{e}-folding radius is $r_e = 174 \pm 18$ kpc, consistent with the value at lower frequencies within errors. For the candidate relic, we measure $S_{\rm 1.4GHz} = 5.2 \pm 0.3$ mJy, leading to $\alpha_{\rm 144MHz}^{\rm 1.4GHz} = -1.2 \pm 0.2$. 

\subsection{Spectral analysis}
\label{sec:spindex}

To investigate the nature of the diffuse emission, we have produced spectral index maps making use the frequency coverage available for A1550. First, HBA, LBA and uGMRT maps were produced by applying the same visibility cut of 80 $\lambda$-14 k$\lambda$ and {\ttfamily Briggs -0.3}, in order to compensate for the different \textit{uv}-coverage of the instruments and match spatial scales. To avoid possible artefacts produced by the source subtraction, this procedure was applied to the non-subtracted data. The spectral index map was generated by estimating $\alpha$ and $\Delta\alpha$ in each pixel as:

\begin{equation}
\label{eq:spindex}
\alpha_{\nu1}^{\nu2} = \dfrac{\ln S_1 - \ln S_2}{\ln \nu_2 - \ln \nu_1} \pm \dfrac{1}{\ln \nu_2 - \ln \nu_1} \sqrt{\bigg(\frac{\sigma_1}{S_1}\bigg)^2 + \bigg(\frac{\sigma_2}{S_2}\bigg)^2} ,
\end{equation}

where $S_1$ and $S_2$ are the flux densities at frequencies $\nu_1$ and $\nu_2$, respectively, while $\sigma$ is the corresponding error. Similarly, spectral index maps between 1.4 GHz (VLA) and 400 MHz were produced by applying a baseline cut at 140$\lambda$-14k$\lambda$ and {\ttfamily Briggs -0.3}. Visibilities were tapered at 30\arcsec\, to highlight the diffuse emission.

The spectral index map between 54 and 144 MHz is shown in the top-left panel of Fig.~\ref{fig:spindexLOFAR}. The spectral index looks flatter (-1$<\alpha<$-0.5) in the position of the AGN and of the point source South to A, as expected from compact objects where self-absorption is relevant. A similar behaviour is observed in the patch labeled B, West to the candidate relic, with the spectrum being flatter ($\alpha \sim$ -0.4) in the centre and becoming steeper ($\alpha \sim$ -0.8) moving to the periphery. This suggests that source B is likely a double-lobed radio galaxy, even though we cannot find any obvious optical counterpart. The region of the halo East to the head-tail shows a steep index ranging between -1.2 and -2.2, which becomes even steeper at the periphery and, more interestingly, West to the tail of the AGN, where source A is located. In this region we observe a mean index of $\alpha \simeq -1.7$, but it reaches $\alpha \simeq -2.3$ around A. The bridge East to the candidate relic also shows a similar spectrum. At the position of the relic, we observe a spectral index which steepens from $\alpha \simeq -0.5$ to $\alpha \simeq -1.8$ in the westernmost region, even though the relatively low resolution makes it hard to observe the typical gradients observed in relics. Finally, the NE extension shows $\alpha \simeq -1.8$ in the direction of the cluster centre, while moving to the outskirts we observe a steepening to $\alpha \simeq -2.2$. 

In the top-middle panel of Fig.~\ref{fig:spindexLOFAR} we show the spectral index map between 144 and 400 MHz. We placed upper and lower limits on the spectral index where the emission was not detected at the higher or lower frequency, respectively. This might be due to the different frequency range of the instruments, and to the spectral curvature of the sources. We observe a number of lower limits around the main emission, which are likely uGMRT calibration artifacts which could not be improved. The region of the halo looks consistent with what observed between 54 and 144 MHz, with $\alpha$ ranging between -1.2 and -1.9. Due to the lower resolution set to detect the halo, it is hard to assess the spectral index gradient in the relic. However, we observe a  steep ($\alpha < -2$) spectrum for source A, for which we are only able to place upper limits because of the low emission detected by uGMRT. This indicates a relatively high curvature at high frequencies, suggesting a cutoff in the electron energy spectrum.

Finally, in the top-right panel of Fig.~\ref{fig:spindexLOFAR} we show the spectral index map between 400 MHz and 1.4 GHz. Upper limits are placed where no emission is observed with VLA. Due to the lower resolution, it is hard to resolve all the structures which are observed by LOFAR (especially HBA) and uGMRT. The spectral index across the halo is consistent with what we found at lower frequencies, and steepens towards the candidate relic and source A, which is not detected at 1.4 GHz. Hints of upper limits $\alpha < -2$ can be observed around the position of source A, even though they are not as clear as in the LOFAR-uGMRT map. In the putative relic we seemingly observe a flattening of the spectrum moving towards the cluster outskirts, even though the low resolution makes this result uncertain.

\subsection{Diffuse radio emission and thermal plasma}
\label{sec:xray}

The short exposure time of the \textit{Chandra} observation does not allow us to perform a thorough analysis of the ICM in A1550. Still, we can study the morphology of the X-ray emission and compare it to the structures observed in the radio band.

\begin{figure}
	\centering
	\includegraphics[scale=0.43]{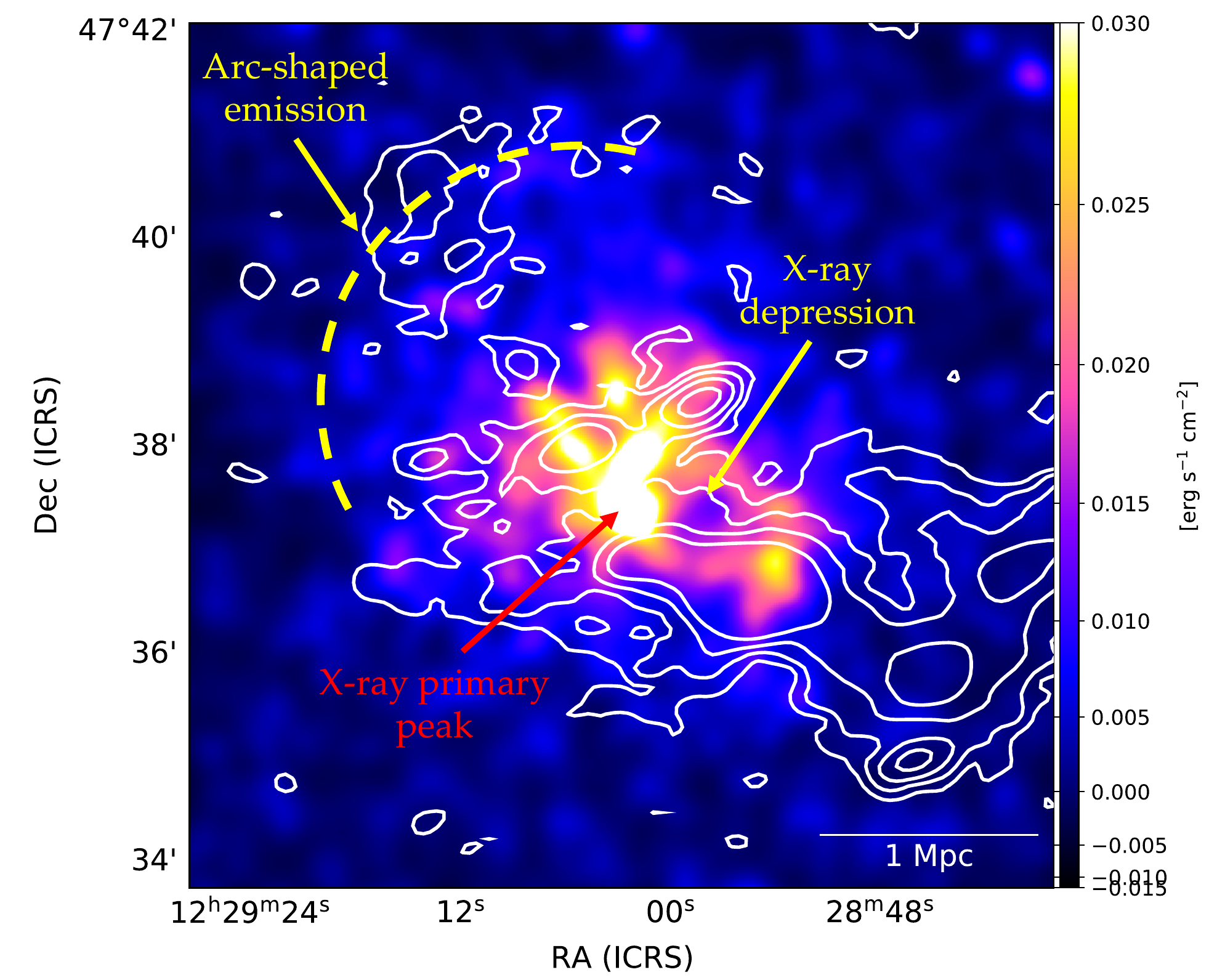}
	\caption{\textit{Chandra} exposure-corrected, background-subtracted 0.5-7 keV image of A1550, smoothed to a resolution of 30 kpc at the cluster redshift. LOFAR LBA contours are overlaid in white. The image is cut at the edge of the CCD FoV.}
	\label{fig:chandra}
\end{figure}

\begin{figure*}
	\centering
	\includegraphics[scale=0.364]{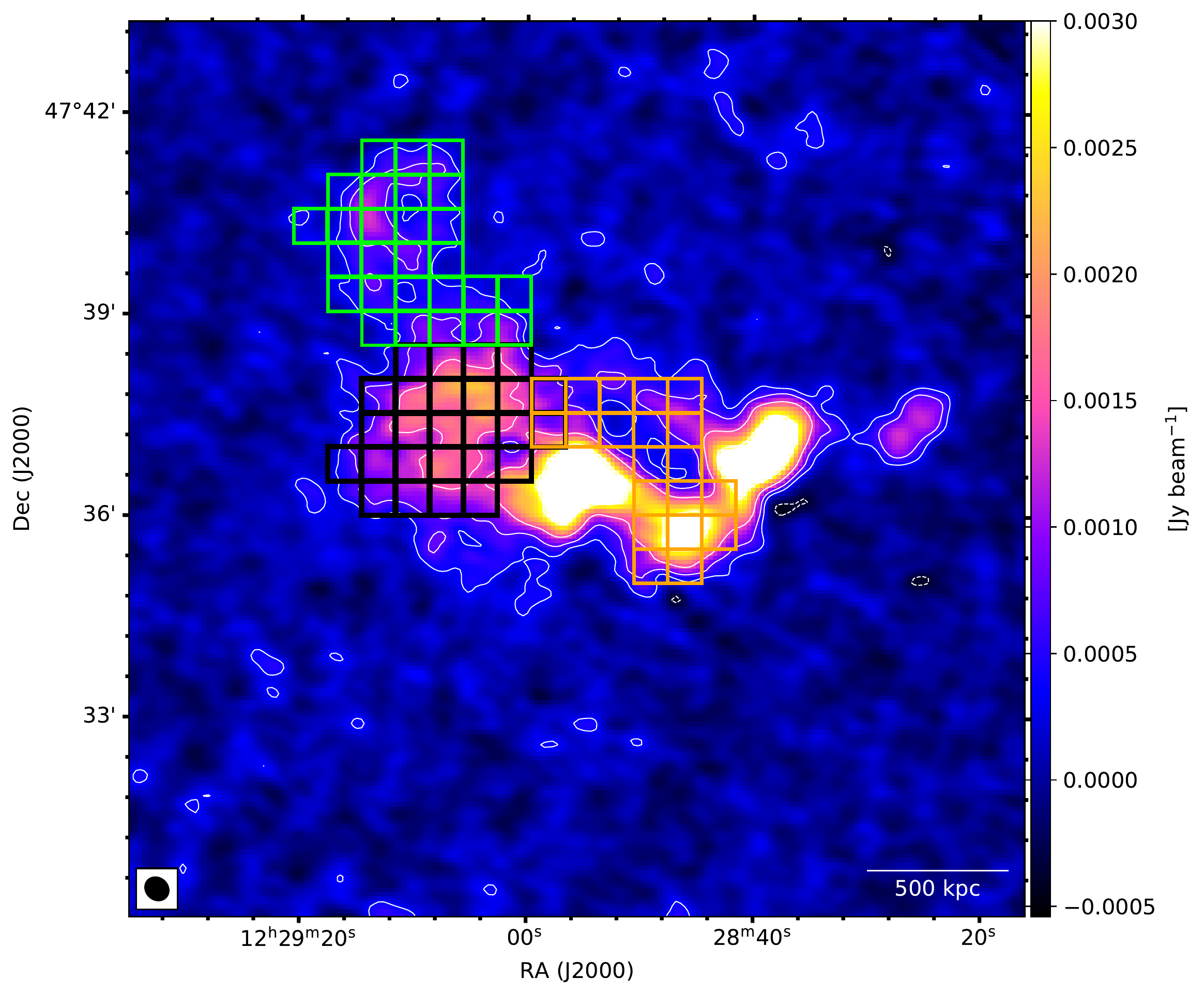}
	\includegraphics[scale=0.36]{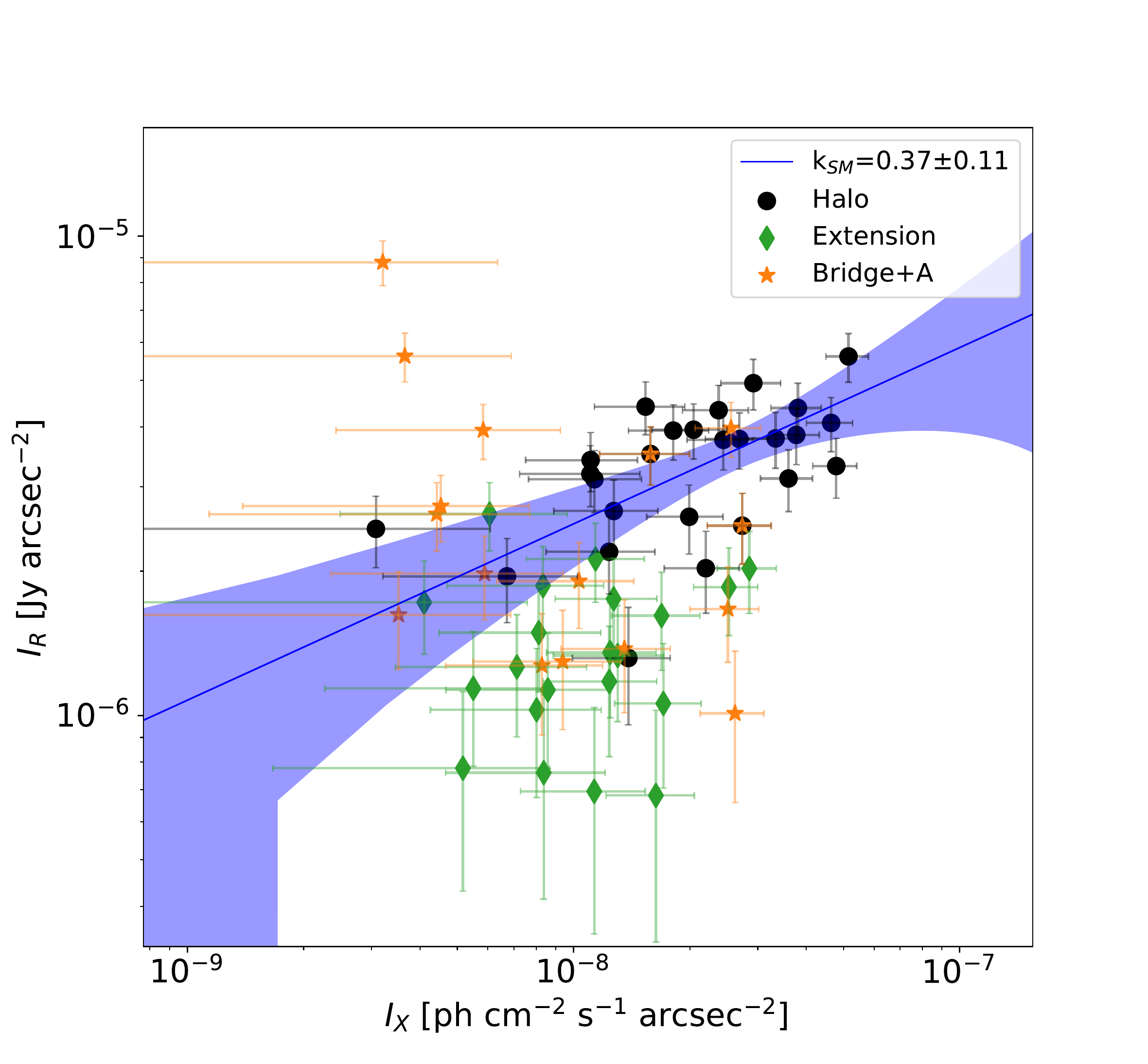}
	\caption{\textit{Left}: Grids used to sample the halo (black), the NE extension (red) and the bridge+source A region (orange). The dimension of the squares is 1.5 times the radio beam, with the beam being 22\arcsec$\times$19\arcsec. \textit{Right}: Point-to-point correlation between radio (144 MHz) and X-ray surface brightness. Black data represents the halo, red represents the NE extension, while orange refers to the South-West region including the bridge and source A.}
	\label{fig:ptp}
\end{figure*}

In Fig.~\ref{fig:chandra} we show the \textit{Chandra} image of A1550 in the 0.5-7 keV band, where the emission of the ICM is best visible. The morphology of the ICM looks roughly elliptical (ellipticity\footnote{Estimated as the ratio between the minor and major axes.} = $\epsilon \sim 0.75$) from the current X-ray image. The morphological parameters presented in Table A.2 of \citealt{Botteon_2022}, such as concentration ($c = (8.67 \pm 0.89) \times 10^{-2}$) and centroid shift ($w = (3.80 \pm 0.35) \times 10^{-2}$), confirm that the cluster is disturbed, accordingly to the thresholds defined in \citet{Cassano_2010}. This is consistent with the presence of radio diffuse emission. The X-ray peak is found at $\sim$150 kpc ($\sim$38\arcsec) from the AGN, and the morphology of the emission seems to roughly correlate with that of the radio halo. The archival observation does not show hints of either X-ray cavities or cold fronts, apart from a $\sim50$ kpc depression between the AGN and its tail. However, due to the significant smoothing that was applied, it is currently hard to confirm whether this structure is real, or just an artefact. This putative detection needs to be supported by a surface brightness analysis to quantify the possible depression, which is not feasible with the current \textit{Chandra} observation. We do not detect X-ray emission in the region of the candidate relic and source A, even though the small area of the CCD does not allow us to cover the whole extent of the radio structure.
A faint, arc-shaped patch of X-ray plasma stretches in the NE direction. Interestingly, the NE radio extension seems to follow the same morphology, and may be confined by the X-ray emission. This suggests that it could be part of the radio halo, as also supported by our HBA and LBA observations which clearly show that the extension is not detached from the halo emission. Nevertheless, this could also be due to projection effects. 
The correlation between the X-ray and radio surface brightness can be of help here to investigate which structures are part of the halo, and which constitute a separate kind of diffuse emission \citep{Bruno_2021, Rajpurohit_2021, Duchesne_2021}. To this end we performed a point-to-point (ptp) analysis, in which the surface brightness at the two bands was sampled through a grid. One of the first applications of this analysis in galaxy clusters can be found in \citet{Govoni_2001}, and a dedicated algorithm was recently published by \citet{Ignesti_2022}. First, we drew three grids: one on the halo, one on the NE extension and one on the bridge and source A (see Fig.~\ref{fig:ptp}). The size of the squares was chosen to be 1.5 times the radio beam. This is large enough to allow for relatively small error bars, especially for the X-ray image where the exposure time is short, but small enough to provide enough statistics. Nevertheless, the available X-ray observation prevents us from using smaller areas, which would increase our statistics.

The surface brightness was then extracted from each square through the procedure extensively described in \citet{Ignesti_2022}. The analysis was carried out on the HBA image, since the structures are better resolved, on the emission above 3$\sigma$. The result is shown for all three grids in Fig.~\ref{fig:ptp}. A sub-linear correlation is found for the halo, with slope $k = 0.37 \pm 0.11$. The Pearson and Spearman coefficients are 0.53 and 0.56, respectively, suggesting that a correlation may exist, albeit weak. Given that such link is always observed in radio halos (see references above), it is likely that the short exposure time of the X-ray image, which led to the fit having low statistics, could be the reason for the relatively low coefficients. Furthermore, the slope is flatter than the values usually found in halos, which range between $\sim$0.5 and $\sim$0.7 \citep[e.g.,][]{Rajpurohit_2021}. Similar sub-linear slopes were also found in the Bullet cluster \citep{Shimwell_2014} and in A520 \citep{Hoang_2019}, and were interpreted as the halo being in a different evolutionary state with respect to typical halos. Longer X-ray observations in the future may provide smaller error bars and a larger number of bins, leading to a more reliable fit. On the other hand, we see no clear correlation for NE extension and for the SW region. The Pearson and Spearman coefficients are 0.18 and 0.13 for the NE extension, and -0.3 and -0.4 for the bridge and source A. The distribution of their X-ray and radio surface brightness values across the plot looks random, suggesting that they might not be part of the halo, but may constitute another kind of diffuse emission. This is relevant especially for the NE extension, given that both LBA and HBA detected patches of emission connecting it to the halo. 

Finally, it would be interesting to perform the same kind of analysis at all four different frequencies, to assess if and how the slope eventually changes. However, as discussed above, the short X-ray exposure would significantly affect the results. Possibly, an accurate comparison of the X-ray and radio morphology in A1550 will be performed in the next future thanks to deeper observations.

%%%%%%%%%%%%%%%%%%%%%%%%%%%%%%%%%%%%%%%%%%%%%%%%%%%%%%%%%%%%%%%%%%%%%%%%%%%%%%

\section{Discussion}
\label{sec:discussion}

There are a number of interesting sources in A1550. As often observed in disturbed galaxy clusters, the optical BCG which hosts the head-tail radio galaxy is found relatively far from the X-ray peak ($\sim$ 150 kpc), suggesting that the hot gas is disturbed. This is also supported by the lack of a central peak in the ICM, as well as from morphological parameters estimated in \citet{Botteon_2022} and already discussed above. The radio diffuse emission spans a length of $\sim$2 Mpc, suggesting that a major merger might have occurred. Optical images are sometimes useful to assess the merger dynamics, but in the case of A1550 we do not see obvious hints from SDSS and PanSTARSS observations, since we do not detect any evidence for a companion cluster. As also briefly discussed in VW21, SDSS detects a subgroup of $\sim$ 20 galaxies North of the extension, which shows the same redshift of A1550. However, it lies far ($\sim$1 Mpc) from the cluster centre and from the diffuse emission. Therefore, given also its relatively small extent, it is unlikely to be the cause of the extended emission detected in A1550. The orientation of the candidate radio relic might suggest that the merger axis lies in the NE-SW direction. This is also supported by the X-ray surface brightness distribution that is elongated in the same direction. Nevertheless, it is possible that diffuse emission is elongated along the line of sight, and that we are therefore underestimating its size. A method to infer the orientation of a cluster through observations of relics was recently suggested in \citet{Wittor_2021}. It exploits the ratio of the total projected X-ray luminosity of the cluster to the projected X-ray luminosity emitted within the candidate relic region. If the ratio approaches $\sim 1$, it is likely that we are observing the cluster merger face-on. On the other hand, lower ratios indicate that the cluster is elongated in the plane of the sky. Unfortunately, the \textit{Chandra} observation does not fully cover the relic, preventing us from performing this test.

\subsection{The ultra-steep radio halo and the NE extension}

Regardless of the cluster orientation, the spectral index observed for the halo at all available frequencies suggests that it is a USSRH. Despite the number of detected USSRH is still low, radio halos with steep indices are being discovered more and more frequently in the last years thanks to the improved observational capabilities of low-frequency instruments such as GMRT, MWA (Murchison Widefield Array) and LOFAR \citep{Shimwell_2016, Wilber_2018, Bruno_2021, diGennaro_2021, Duchesne_2022}. An in-depth analysis of all radio halos hosted in Planck clusters and observed in LoTSS, including A1550, has recently been presented in \citet{Botteon_2022}. USSRH are a prediction of turbulent re-acceleration models \citep{Cassano_2006, Brunetti_2008}, in which particles are re-accelerated by turbulence \citep{Brunetti_2001, Petrosian_2001, Brunetti_2011, Brunetti_2017}. On the other hand, the detection of such steep indices is not expected from hadronic (or secondary) models, in which the emission of halos comes from the production of secondary electrons from hadronic collisions between thermal and CR protons \citep{Blasi_1999, Dolag_2000, Pfrommer_2008}.
Given that the integrated spectral index observed for the USSRH with LOFAR is $\alpha_{54 \rm MHz}^{144 \rm MHz} \sim -1.6$, we expect an index for the spectral energy distribution\footnote{Defined as $N(E) \propto E^{-\delta}$, with $E$ being the energy.} $\delta = 2\alpha -1 =-4.2$. If there is no break in the spectrum, the energy budget for these particles would be untenable \citep{Brunetti_2008}. Therefore, a break at low energies ($\sim$ GeV) should exist, suggesting a possible interplay between radiative losses and turbulent re-acceleration during the lifetime of emitting electrons \citep{Brunetti_2014}.
Moreover, re-acceleration models predict that a large fraction of halos associated with clusters of masses between 4-7 $\times 10^{14}$ M$_\odot$ should exhibit steep spectra \citep{Cassano_2010, Cassano_2012, Brunetti_2014, Cuciti_2021}. The mass of A1550 of $\sim 6 \times 10^{14}$ M$_\odot$ estimated from \citet{Planck_2016} falls in this range.\footnote{Since the correlation between the halo luminosity and the cluster mass has recently been studied for the whole LOFAR Planck-DR2 sample in \citet{Botteon_2022}, including A1550, we will not report it again in this work.}

Assuming a NE-SW merger axis, the morphology and orientation of the NE extension suggest that this might also be a candidate relic. It is symmetric to the SW candidate relic with respect to the X-ray peak, and both relics lie along the merger axis. It is possible that the extension was generated by the re-acceleration of electrons caused by the counter-shock of the front that produced the SW relic. However, the X-ray observation is too shallow and the \textit{Chandra} FoV too limited to detect any shock, in both directions. Furthermore, the spectral index distribution looks rather uniform across the source, albeit it steepens in the Easternmost part. In fact, the same spectral index distribution seems more consistent with the halo, and even slightly steeper in some regions. Low-resolution images at both LOFAR frequencies (see e.g. Fig.~\ref{fig:subtractedLOFAR}) show a patch of emission connecting the extension to the halo. The \textit{Chandra} observation also detects X-ray plasma in the same region which seems to confine the radio emission. This supports the idea that the extension is part of the central radio halo. If this were true, one would expect the point-to-point analysis for the halo and the extension to have a similar brightness distribution. This is apparently not the case (see Fig.~\ref{fig:ptp} and Sec.~\ref{sec:xray}), even though the X-ray exposure time is too short to allow for good statistics. As for the bridge, deeper X-ray observations could shed light on the nature of the detected NE emission. Similar structures have also been detected in other radio halos \citep{Markevitch_2010, Bonafede_2022, Hoang_2019}, and are presumably due to advection of plasma by motions and shocks.
In this case, the compression of magnetic fields in such structure and the abrupt drop of the brightness upstream of the edge may also affect the ptp correlation, which could explain why we do not see the same trend observed for the halo.

\subsection{The relic and the bridge}

\begin{figure*}
	\centering
	\includegraphics[scale=0.65]{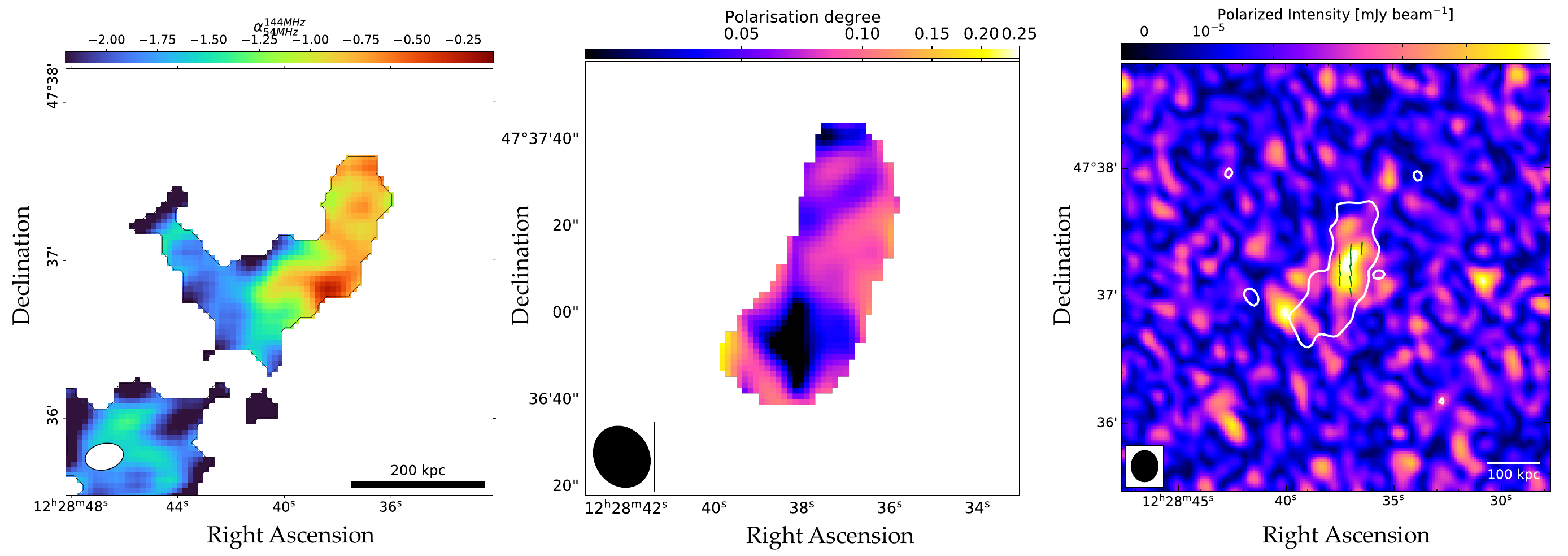}
	\caption{\textit{Left}: High-resolution spectral index map of the candidate relic between 54 and 144 MHz. The beam (15\arcsec$\times$10\arcsec) is shown on the bottom-left as a white circle. Only pixels with flux density above 4$\sigma$ are shown. \textit{Middle}: Polarisation degree map for the relic in A1550. \textit{Right}: Polarised intensity map of A1550. Magnetic field vectors are overlaid in green, while 3$\sigma$ total power contours of the relic are shown in white. The vectors are corrected for Faraday rotation effect.}
	\label{fig:highspectral}
\end{figure*}

 The source that we have classified as a relic shows an elongated shape and lies at $\sim$1 Mpc from the X-ray primary peak. However, due to the relatively low resolution of the LOFAR spectral index map (left panel of Fig.~ \ref{fig:spindexLOFAR}) it is hard to assess whether there is a real spectral index gradient towards the cluster center. We have produced a high-resolution (15\arcsec$\times$10\arcsec) spectral index map between 54 and 144 MHz (shown in Fig.~\ref{fig:highspectral}) since these frequencies provide the best combination of resolution, sensitivity and \textit{uv}-coverage.

In the easternmost part, close to the bridge, the spectral index is as steep as $\alpha \sim -2$. Then, moving towards the outskirts, we observe a flattening from $\sim -1.5$ to $\sim -0.75$, reaching a lower limit of $\sim -0.2$ at the edge. The flatter spectrum is aligned with the presumed orientation of the merger, suggesting that the shock front has first accelerated electrons closer to the cluster centre, and is now travelling in the NE-SW direction. Unfortunately, we  cannot independently confirm the presence of a shock at this location via X-ray observations. 

We have produced polarisation maps for the candidate relic, as described in Sec. \ref{sec:vla}. As shown in Fig.~\ref{fig:highspectral} middle panel, a large portion of the relic shows a degree of polarisation which varies from $\sim10\%$ to $\sim25\%$. Interestingly, the highest polarisation is found in the South-Eastern region, close to the bridge and source A. The magnetic field vectors are aligned with the source orientation (see right panel of Fig.~\ref{fig:highspectral}). The combination of elongated shape, spectral index gradient, location and degree of polarisation confirm the classification as a relic.

As already discussed in Sec.~\ref{sec:results}, we measure a total extent of $\sim$500 kpc at LOFAR frequencies and $\sim$460 kpc at 1.4 GHz. It is well-known that the radio power of relics correlates with the relic LLS \citep{Bonafede_2012}, as well as with the host cluster mass \citep{deGasperin_2014}. To investigate whether the dynamic environment of A1550 has affected such relations, we can compare the radio power, LLS and cluster mass with systems studied in \citet{Bonafede_2012}, for the radio power-LLS correlation, and \citet{deGasperin_2014} for the radio power-cluster mass correlation. We find that the properties of the relic in A1550 are consistent with both relations.

We see that the relic is connected to the halo via a bridge. Similar cases of connected diffuse emission were also detected in \citet{Markevitch_2005}, \citet{Markevitch_2010}, in the Toothbrush cluster \citep{vanWeeren_2012b, Rajpurohit_2018, deGasperin_2020}, in A3667 \citep{Carretti_2013, deGasperin_2022} and others \citep[e.g.,][]{Bonafede_2018}. In most of these clusters the halo is directly connected to the relic, at least seen in projection. One of the most spectacular examples of bridges is found in the Coma cluster where turbulence is believed to produce the radio emission in the bridge \citep{Bonafede_2021}. In A1550 the nature of sources is different from Coma, since the bridge is apparently not related to the radio galaxy. If the merger proceeded along the NE-SW axis, as suggested by the (projected) orientation of the relic and by the morphology of the X-ray emission, then the bridge could consist of electrons that were pre-accelerated by the shock (e.g., \citealt{Fujita_2016}), and then re-accelerated by the substantial merger-driven turbulence (e.g., \citealt{Gaspari_2014}). However, the source of seed electrons is unclear. Furthermore, an elliptical region void of radio emission, labelled as 'depletion' in Fig. \ref{fig:overlay} and with a major axis of $\sim$ 400 kpc, is observed North of the radio galaxy. It is confined by the halo in the East and the relic and the phoenix in the West. It is possible that this region is related to a low-density region in the ICM since a spherical depression is also observed in the \textit{Chandra} image (see Fig.~\ref{fig:chandra}). It is currently not clear if and how this depletion can be physically connected to the bridge. 

\subsection{Source A and gentle re-acceleration}
\label{sec:gentle}

\begin{figure*}
	\centering
	\includegraphics[scale=0.35]{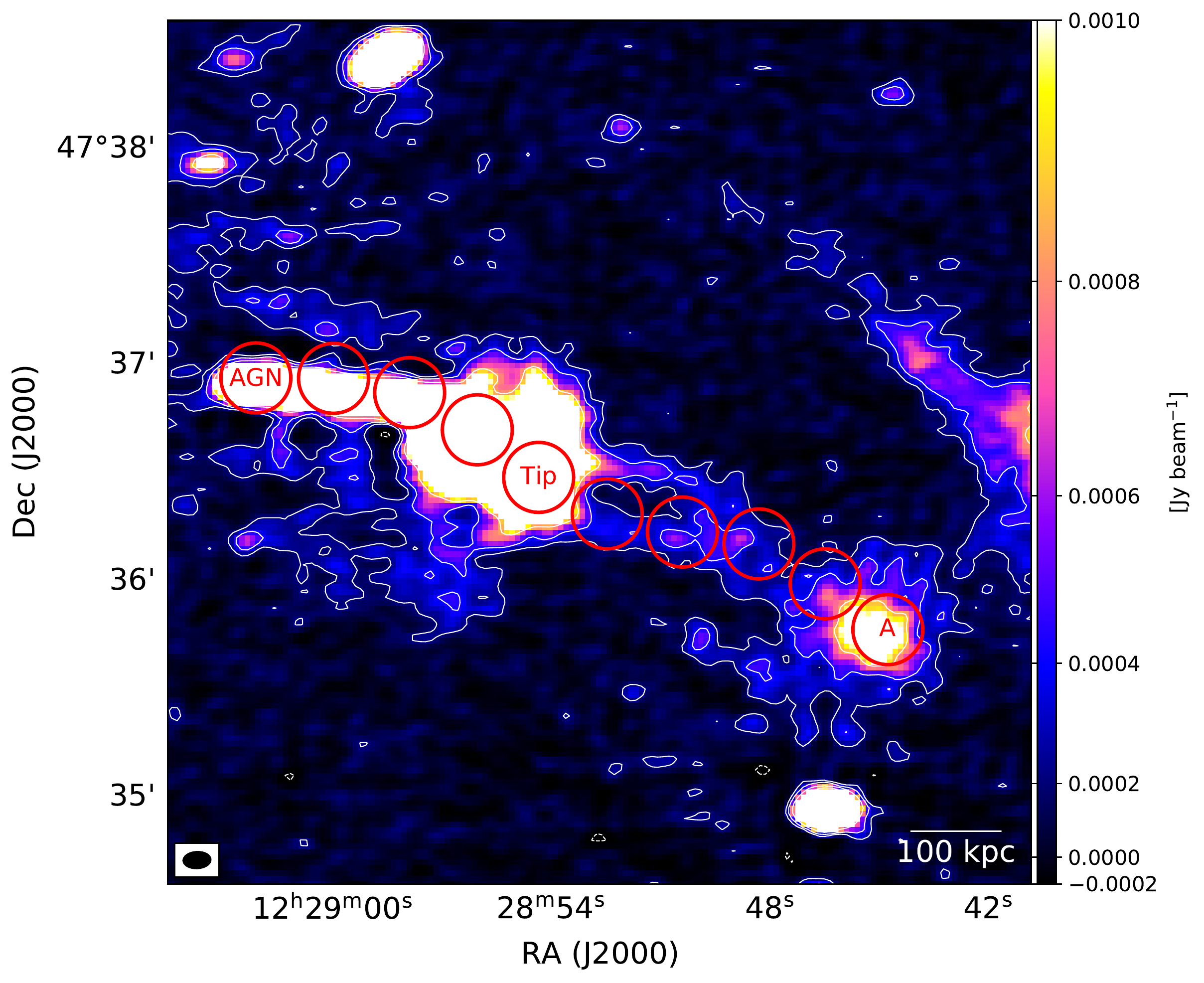}
	\includegraphics[scale=0.63]{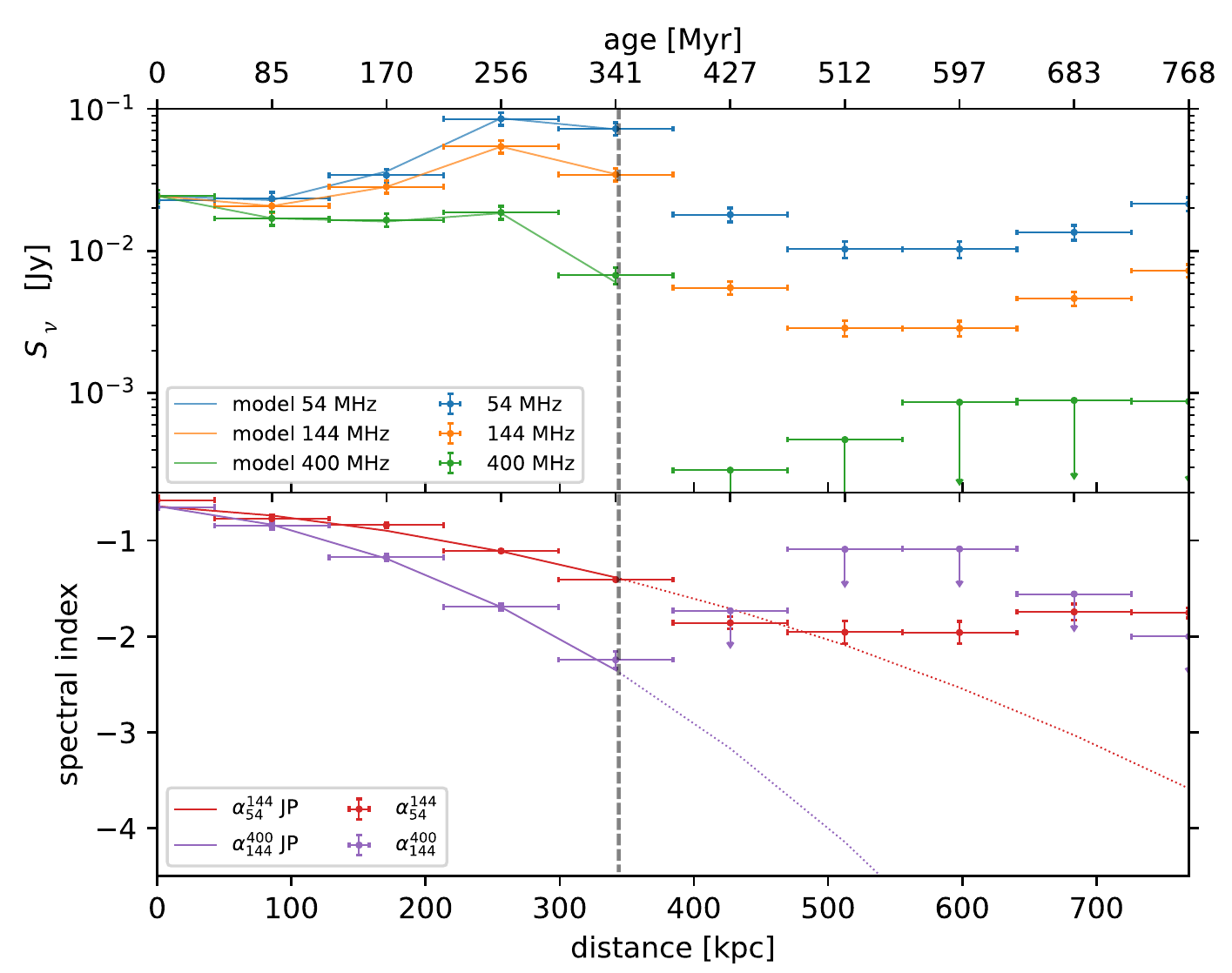}
	\caption{\textit{Left}: Regions from which the spectral index between 54 and 144 MHz was extracted, overlaid on the high-resolution HBA map. Each region is as large as the smallest common circular beam (9.7\arcsec $\times$ 9.7\arcsec). The AGN, the tip of the AGN tail and source A are labeled. \textit{Right}: Results of the fit of the spectral index. Top panel:  flux density at 54 (blue) and 144 (orange) MHz for each region, starting from the AGN position. Bottom panel:  spectral index, fitted with a JP aging model. The curve becomes dashed from the point where the model was extrapolated, indicated by the grey vertical line.}
	\label{fig:samplespindex}
\end{figure*}

In the HBA high-resolution map (see Fig.~\ref{fig:br0LOFAR}) we observe two faint arms\footnote{Also visible in Fig.~\ref{fig:samplespindex}.} that depart from the bright tail of the central AGN, connecting the centre of the diffuse emission to source A, and that might be related to previous outbursts of the AGN. Although the tail of the radio galaxy looks well-confined, the high-resolution image seems to suggest that the electrons which constitute the arms are originally injected from the head-tail. Corresponding to source A, the surface brightness increases again, producing a roundish structure with no visible counterparts in optical (SDSS) catalogs and spectral index $\alpha_{54 \rm MHz}^{144 \rm MHz} = -1.9 \pm 0.2$. When moving to 400 MHz, from the spectral index map in the middle panel of Fig.~\ref{fig:spindexLOFAR} we find an upper limit for the index of $\alpha_{144 \rm MHz}^{400 \rm MHz} \leq -2$. This hints at a curved spectrum, as also shown in Fig.~\ref{fig:spectra}\footnote{The flux density for source A at 1.4 GHz is an upper limit.}. The increase in terms of surface brightness compared to the region closer to the AGN, combined with the spectrum break-off, suggests that source A could be AGN plasma (e.g. phoenix) which was revived through re-acceleration. To trace the behaviour of the electron energy spectrum from the tail to source A, we have sampled the spectral index in regions as large as the smallest common beam (9.7\arcsec $\times$ 9.7\arcsec) from high-resolution HBA, LBA and uGMRT images, which allow us to resolve both the arms and source A, and to get a reasonable statistics. This is done with the assumption that the arms are related to the head-tail radio galaxy, even though it is possible that they might just constitute some form of bridge/filaments, as observed East of the relic. We fit the distribution by applying a non-linear least squares regression, with a Jaffe-Perola (JP) electron aging model \citep{Jaffe&Perola_1973}, which is assumed to be the same (i.e. same break frequency at same distance) at all frequencies. According to that model, the ageing  only depends on the magnetic field and the projected velocity of the radio galaxy through the ICM. We have assumed a minimum energy magnetic field of $B_{\rm min} = \frac{3.25}{\sqrt 3} \times 10^{-10} (1+z)^2$ T = 2.95 $\mu$G, similar to Edler et al. (subm.). We fit the model directly to the spectral index, with the only free parameter being the radio galaxy velocity, to remove any dependence on the normalisation (see Edler et al. subm.). The sampling regions were drawn from the high-resolution HBA image, and are shown in Fig.~\ref{fig:samplespindex} together with the results of this analysis.

The spectral index between 54 and 144 MHz from the AGN to the tip of the tail, at a distance of $\sim$350 kpc, is in the range $-0.58 < \alpha_{54 \rm MHz}^{144 \rm MHz} < -1.41$. From the AGN position to the tip, the index keeps getting steeper as we are moving further from where electrons are injected. From the tip to source A, the distribution reaches a plateau around $\alpha_{54 \rm MHz}^{144 \rm MHz} \sim -1.9$, and then finally flattens at $\alpha_{54 \rm MHz}^{144 \rm MHz} \sim -1.7$ in the region of source A. The fit with a JP aging model is accurate ($\chi^2/DoF \sim 0.8$) along the tail. However, when we move towards the arms and source A, it seems that the spectral index remains fairly uniform and does not follow a pure aging model. This hints at some kind of re-acceleration process which re-energized the electrons in the region of source A. An alternative explanation is that the flatter regions are produced by an increase in the magnetic field strength. Nevertheless, this would require a precise fine-tuning of the magnetic field, making this solution less likely.

A possibility is that old radio plasma from past AGN outbursts was compressed by the same shock that produced the relic. The shock passing through fossil plasma could have produced the steep, curved spectrum (see \citealt{Ensslin_2001} and Sec.~\ref{sec:intro} for more references). The curvature observed at 400 MHz for source A from Fig.~\ref{fig:spindexLOFAR} and the (likely) location of the shock front support this hypothesis. The small size ($\sim$130 kpc radius) of this source is also consistent with expectations from revived plasma (VW21), as well as the apparently irregular distribution of the spectral index observed around source A \citep{Kale_2012}.

Finally, the plateau shown by the spectral index distribution in the region of the two arms might point to some kind of gentle re-acceleration, which is barely able to compensate the radiative losses of the electrons. A similar case has been studied in \citet{deGasperin_2017} in the galaxy cluster A1033, where the spectral index distribution across a long ($\sim$500 kpc) radio galaxy tail remains level with increasing distance from the injection point (i.e. the AGN). The re-acceleration that we see in A1550 might be similar, even though the plateau in A1033 is reached at much steeper spectral indices ($\alpha \sim -4$), at least between 144 and 323 MHz. In A1033, compression by shocks is unlikely to explain the plateau as this would require precise geometrical tuning: the tail is so long that it would be hard for a shock to re-accelerate all electrons at once. Furthermore, simulations have shown that even mild (Mach number $\mathcal{M} < 2$) shocks are able to disrupt radio galaxy tails \citep[e.g.][]{O'Neil_2019, Nolting_2019}. In A1550 the two arms are shorter than the tail in A1033, but it remains unlikely for a shock to compress and re-accelerate all particles at the same time across 300 kpc in order to produce a uniform spectral index (Fig.~\ref{fig:samplespindex}). Therefore, it is possible that the same gentle re-energisation which was invoked for A1033 can explain  the two arms of A1550.

%%%%%%%%%%%%%%%%%%%%%%%%%%%%%%%%%%%%%%%%%%%%%%%%%%%%%%%%%%%%%%%%%%%%%%%%%%%%%%

\section{Conclusions}
\label{sec:conclusions}

We have studied a plethora of diffuse emission sources in the galaxy cluster A1550 through multi-frequency radio observations, ranging from 54 MHz to 1.4 GHz. Our results can be summarised as follows:

\begin{itemize}

    \item We observe an ultra-steep spectrum radio halo with a mean spectral index $\alpha \sim -1.6$, and an extent of $\sim$ 1.2 Mpc at 54 MHz. The halo encompasses the head-tail radio galaxy at the centre of the cluster, and extends towards the East. The detection of a USSRH favours primary models for the origin of the relativistic electrons.
    
    \item We found a relic West of the head-tail radio galaxy, with a projected extent of $\sim$ 500 kpc. From the relic, a bridge departs towards East and connects it to the halo, similarly to other clusters (see Sec.~\ref{sec:discussion} for further details).
    The X-ray emission morphology and the orientation of the diffuse emission, and especially of the relic, point to a merger which occurred in the NE-SW direction. Then, the bridge could consist of electrons that were pre-accelerated by shocks, and then re-accelerated by the turbulence which also produced the halo.
    
    \item East of the halo, we detect an extension at low-frequencies that appears to depart from the halo. Even though its shape and orientation resembles a relic, the spectral index distribution indicates that it is an extension of the halo, as also suggested by the physical connection observed between the two. The point-to-point analysis performed on radio and (short exposure) X-ray images does not provide conclusive results.
    
    \item Between the radio galaxy and the relic, we observe a roundish source that we classify as a radio phoenix. This is supported by the steep and curved spectrum, which shows $\alpha \sim -1.9$ at LOFAR wavelengths, but steepens at values $\alpha < -2.1$ moving to higher frequencies. These electrons may have been re-accelerated through adiabatic compression by the same shock that has produced the relic.
    
    \item Two arms connect the tail of the radio galaxy to the phoenix. Our analysis shows that the spectral index in the arms does not steepen with increasing distance from the electron injection point. Instead, it reaches a plateau around $\alpha \sim -2$, which is maintained for more than 300 kpc. This is similar to the mild/gentle re-acceleration observed in GReET sources. It is also possible that the same shock which produced the relic and the phoenix was able to re-accelerate electrons along the arms, leading to the observed plateau in the spectral index distribution. However, it remains difficult for a shock that travels in the direction along the arms,to re-accelerate all electrons across 300 kpc at once.
    
\end{itemize}

Determining the origin of the different sources in the cluster A1550 remains complex. Deeper X-ray observations are needed to identify shocks in the ICM in order to determine the orientation of the merger and study the turbulence that produced the USSRH. A1550 is among the first of the LoLSS-HETDEX clusters which we have studied in detail at very low frequencies. In the future, it will be possible to perform a similar analysis on a much larger sample of objects. The synergy with other radio telescopes, such as GMRT and VLA, and complementary X-ray data will be key to shed more light on re-acceleration processes in galaxy clusters.

\section*{Acknowledgements}\label{acknowledgments}
\scriptsize
TP is supported by the DLR Verbundforschung under grant number 50OR1906. MB acknowledges support from the Deutsche Forschungsgemeinschaft under Germany's Excellence Strategy - EXC 2121 "Quantum Universe" - 390833306. AB acknowledges support from the VIDI research programme with project number 639.042.729, which is financed by the Netherlands Organisation for Scientific Research (NWO). RJvW acknowledges support from the ERC Starting Grant ClusterWeb 804208. FG, GB and MR acknowledge support from INAF main- stream project ‘Galaxy Clusters Science with LOFAR’ 1.05.01.86.05. MG acknowledges partial support by NASA Chandra GO9-20114X and HST GO-15890.020/023-A, and the {\it BlackHoleWeather} program. GDG acknowledges support from the Alexander von Humboldt Foundation. DNH and CJR acknowledges support from the ERC through the grant ERC-Stg DRANOEL n. 714245.
LOFAR \citep{vanHaarlem_2013} is the
LOw Frequency ARray designed and constructed by ASTRON. It has observing, data processing, and data storage facilities in several countries, which are
owned by various parties (each with their own funding sources), and are collectively operated by the ILT foundation under a joint scientific policy. The
ILT resources have benefitted from the following recent major funding sources:
CNRS-INSU, Observatoire de Paris and Université d’Orléans, France; BMBF,
MIWF-NRW, MPG, Germany; Science Foundation Ireland (SFI), Department
of Business, Enterprise and Innovation (DBEI), Ireland; NWO, The Netherlands; The Science and Technology Facilities Council, UK; Ministry of Science and Higher Education, Poland; Istituto Nazionale di Astrofisica (INAF),
Italy. This research made use of the Dutch national e-infrastructure with support
of the SURF Cooperative (e-infra 180169) and the LOFAR e-infra group, and
of the LOFAR-IT computing infrastructure supported and operated by INAF,
and by the Physics Dept. of Turin University (under the agreement with Consorzio Interuniversitario per la Fisica Spaziale) at the C3S Supercomputing Centre, Italy. The Jülich LOFAR Long Term Archive and the German LOFAR
network are both coordinated and operated by the Jülich Supercomputing Centre (JSC), and computing resources on the supercomputer JUWELS at JSC were
provided by the Gauss Centre for Supercomputing e.V. (grant CHTB00) through
the John von Neumann Institute for Computing (NIC). This research made use
of the University of Hertfordshire high-performance computing facility and the
LOFAR-UK computing facility located at the University of Hertfordshire and
supported by STFC [ST/P000096/1].

\bibliographystyle{aa.bst}
\bibliography{bibliography}

\begin{appendix}
\onecolumn

\section{Results of the Halo-FDCA pipeline}
\label{fdca_appendix}

\begin{figure}[h!]
	\centering
	\includegraphics[scale=0.412]{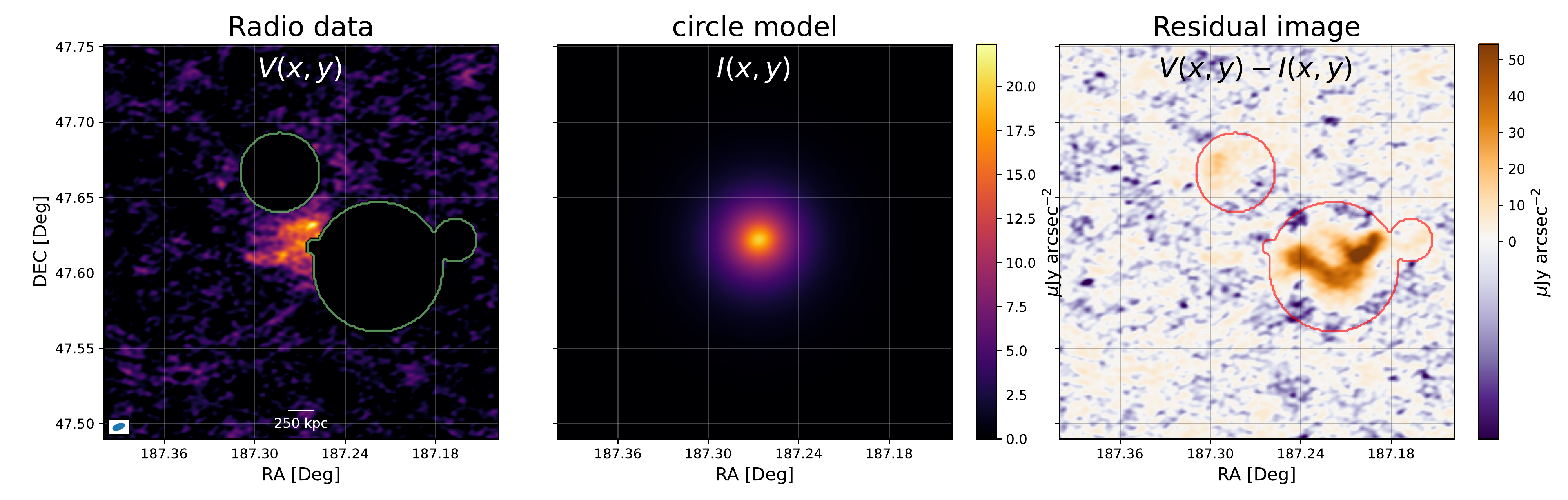}
	\includegraphics[scale=0.412]{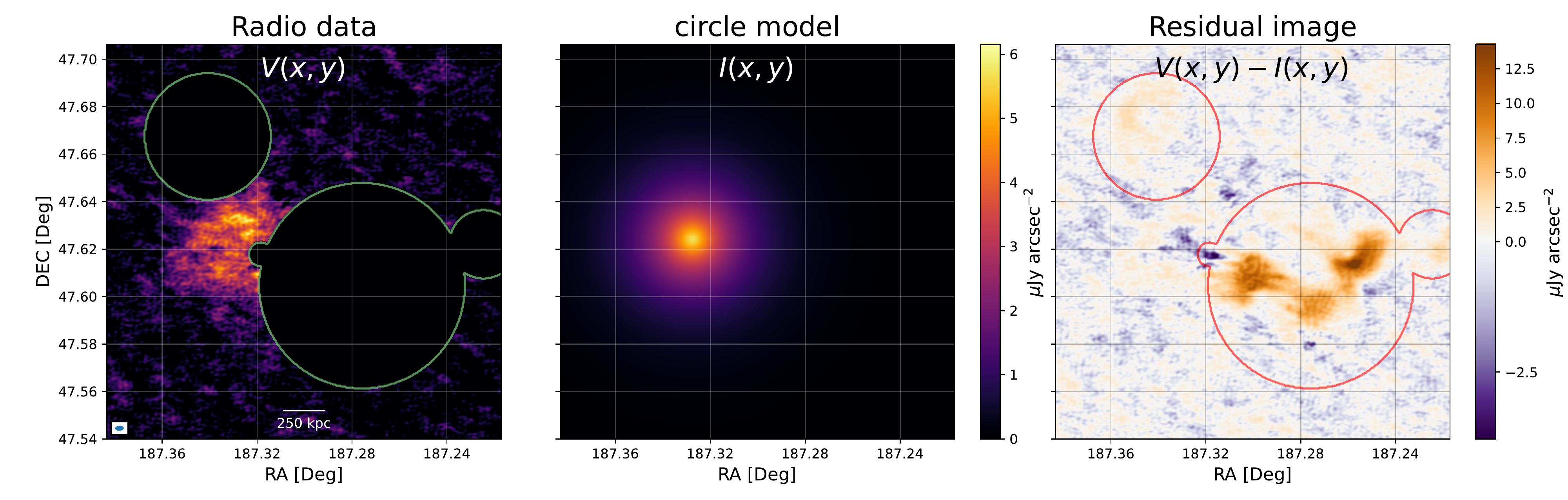}
	\includegraphics[scale=0.412]{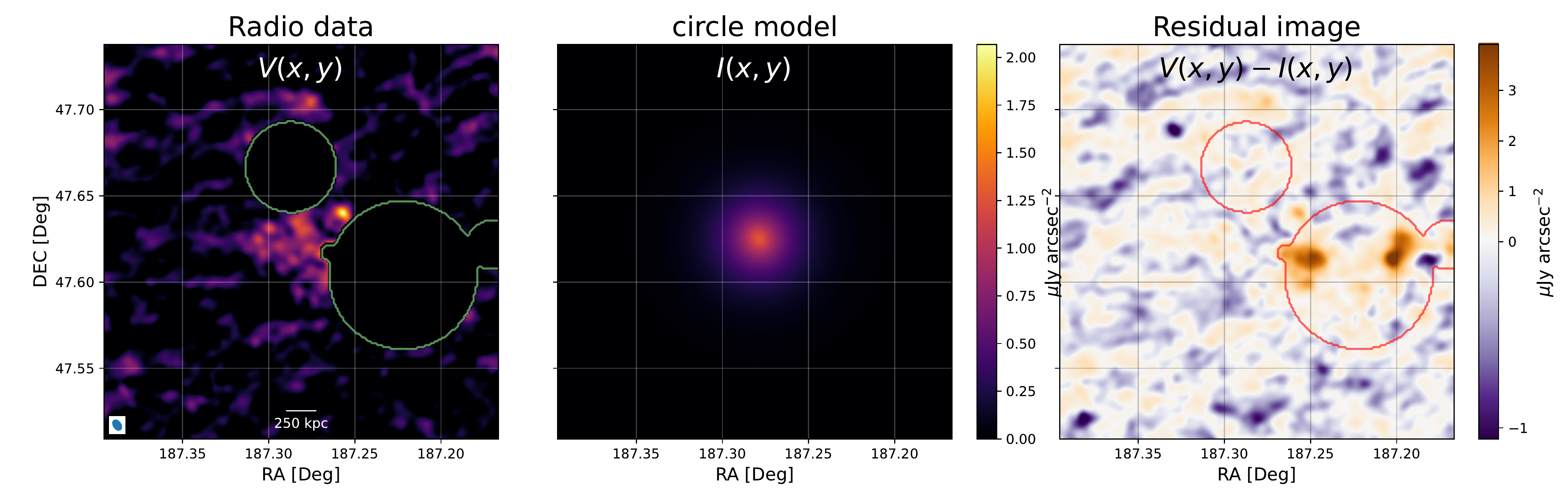}
	\includegraphics[scale=0.412]{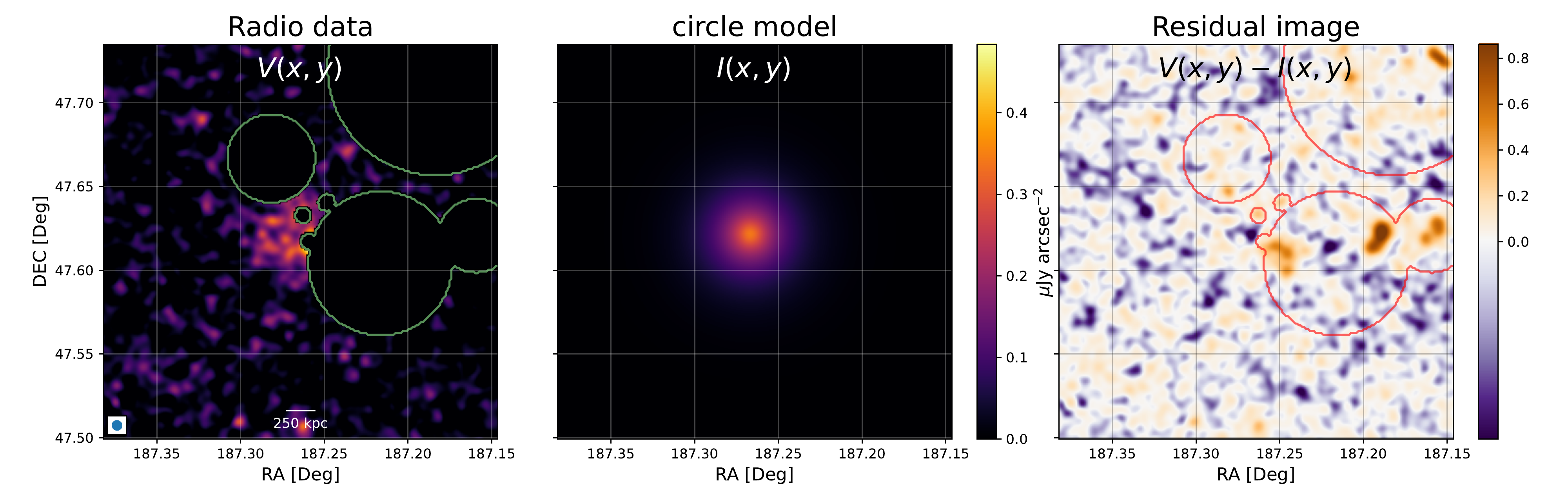}
	\caption{From top to bottom, from left to right: 54, 144, 400 and 1400 MHz modelling of the radio halo, halo centre as detected from a circle model, and residual images with masks applied to exclude sources from the flux density calculation.}
	\label{fig:appendix}
\end{figure}

In Fig.~\ref{fig:appendix} we show the output of the Halo-FDCA pipeline for the estimate of the halo flux density at the different frequencies. We fit a circle model and masked all the emission which we do not classify as being part of the halo. We applied the same masks at all frequencies except for 1.4 GHz, where we had to add more due to a trickier source subtraction (see Sec. \ref{sec:results}).

\end{appendix}

%\bsp	% typesetting comment
\label{lastpage}
\end{document}